\documentstyle[psfig]{article}
\makeatletter
\@addtoreset{equation}{section}

\makeatother

\def\spose#1{\hbox to 0pt{#1\hss}}
\def\simlt{\mathrel{\spose{\lower 3pt\hbox{$\mathchar"218$}}
     \raise 2.0pt\hbox{$\mathchar"13C$}}}
\def\simgt{\mathrel{\spose{\lower 3pt\hbox{$\mathchar"218$}}
     \raise 2.0pt\hbox{$\mathchar"13E$}}}
\def\beq{\begin{equation}}
\def\eeq{\end{equation}}
\def\bea{\begin{eqnarray}}

\def\eea{\end{eqnarray}}
\def\ds{\displaystyle}
\def\nsz{\normalsize}
\def\ssz{\scriptsize}

\def\ni{\noindent}
\def\ul{\underline}
\def\ol#1{\overline{#1}}
\def\req#1{(\ref{#1})}

\def\Tr{\mbox{Tr}}

\def\ln{\mbox{ln}}
\def\Det{\mbox{Det}}

\def\pseudocapt#1{\nopagebreak{\protect\small #1}}
\def\barc{\begin{array}{c}}
\def\ear{\end{array}}

\def\bit{\begin{itemize}}
\def\eit{\end{itemize}}

\title{}
\author{}
\date{}

\begin{document}

\ni{\bf\Large Information content in uniformly discretized Gaussian noise:
optimal compression rates} \\
\ni August Romeo, Enrique Gazta\~naga, Jose Barriga, Emilio Elizalde  \\
\ni Consejo Superior de Investigaciones Cient\'{\i}ficas (CSIC) \\
\ni Institut d'Estudis Espacials de Catalunya (IEEC) \\
\ni Edifici Nexus--201, c. Gran Capit\`{a} 2--4, 08034 Barcelona, Spain\\

\vspace*{.5cm}
Published in: International Jounal of Modern Physics C, Vol.10, 687-716 (1999)
\vspace*{.5cm}

\ni{\bf Abstract.} {\nsz 
We approach the theoretical problem
of compressing a signal dominated by Gaussian noise.
We present expressions for the compression ratio which can be reached, 
under the light of Shannon's noiseless coding theorem, 
for a linearly quantized stochastic Gaussian signal (noise).
The compression ratio decreases logarithmically
with the amplitude of the frequency spectrum $P(f)$ of the noise.
Entropy values and compression
rates are shown to depend on the shape of this power spectrum, 
given different normalizations.
The cases of white noise (w.n.), 
$f^{n_p}$ power-law noise ---including $1/f$ noise---,
(w.n.$+1/f$) noise, and piecewise (w.n.+$1/f |$ w.n.$+1/f^2$) noise
are discussed, 
while quantitative behaviours and useful approximations are provided.

\ni{\bf Keywords:} Information Theory, Signal compression}

\section{Introduction}

There are several motivations to consider the theoretical problem
of compressing noise (or signals so stochastic that deserve this name).
In some cases, the signal to be transmitted is intrinsically noisy 
(e.g. from scientific measurements) and
needs to be compressed in a lossless way before any reduction process 
can be applied.
One of the measured quantities which best exhibits this intrinsic
randomness
is the fluctuation of the cosmic microwave background (CMB) radiation.
Considerable efforts have already been made in order to cope with
the handling of such sort of data (see e.g. \cite{Hi}-\cite{Wa}).
Like other signals from scientific instruments on-board
space satellites, CMB-measurements produce high rates of noisy data
that have to be sent to Earth via a more or less limited telemetry rate
\cite{PhaseA}.

Electronic instruments (e.g. detectors, amplifiers) show characteristic low 
frequency instabilities ($1/f$ noise) to be added to white or thermal noise.
When the signal measured with these instruments is weak, it can only be 
recovered from averaging many measurements. 
The averaging is possible only after a
careful calibration of the low frequency instabilities, which in practice 
means that the whole (noisy) signal has to be transmitted (to Earth).
This is an example that requires lossless compression of a signal dominated 
by noise. In the present work we would like to study, in a quantitavive way,
to what extent noise can be compressed.

This noise is usually treated as a Gaussian stochastic process with an 
arbitrary power spectrum (some relevant aspects of this type of processes have
been considered in \cite{Pa}-\cite{Th}).
We shall assume that its values are discretized 
---{\it quantized}--- in a uniform or linear way.
Given the properties of a Gaussian distribution,
it is possible to find analytical approximations for its information content,
and we will take adavantage of them for obtaining 
the ideal ---i.e., highest theoretically achievable---  compression factor.

In the present work
we make no reference to the error brought about by the discretization
process itself. Yet, a few words on this subject are perhaps called for.
A typical measure of the error caused is the {\it distortion} $D$
(many aspects of rate distortion theory are covered in \cite{Be}. 
The same philosophy
has been applied to the minimum discrimination information (MDI)
theory ---see e.g. \cite{EHG} and refs. therein).
This magnitude
is a sum ---or integral--- of the error
between the continuous values of the initial random variable and the 
associated discrete ones, weighted by the probability distribution.
According to the theory,
for a given initial length of the random variable
there is a minimal possible distorsion.
Then, $\sqrt{D}$ may be interpreted as the lowest `distortion noise'.
Usually, this optimal $D$ falls as the length increases, but
this implies to increase the entropy, thus setting a trade-off between 
compressibility and distortion.

The text is organized as follows.
In Section 2 we present a basic introduction to the problem of 
data compression. 
In Section 3 we deal with the one-dimensional Gaussian case,
which will be helpful for studying multidimensional Gaussian noise with 
possible correlations in Section 4.
Information content and compression values are then discussed.
In Section 5 our conclusions are presented.
A number of calculations have been included in the appendices.

\section{The basic data compression problem}

Standard lossless data compression techniques are applied 
successfully only to data sets with some redundacy.
This redundancy can be formally expressed using the entropy $H$.
It is easy to show (see below) that  it is not possible to
compress a (uniformly) random distribution of  measurements.
If noise is discretized to a high resolution 
(as compared to its variance) the resulting distribution of numbers 
approaches a uniform distribution. 
This indicates that lossless compression might not be very efficient
when the data is dominated by noise, but, as we shall see, the problem depends
crucially on the digital resolution and the range of values to
be stored.

Hypothetical data compression problems can be considered
in the light of Shannon's first theorem (see \cite{Shannon}).
This theorem tells us
that the Shannon entropy $H$ of a source is the lower bound to 
the average length of the code units or `words'
(In addition, we know that such a lower bound can be fairly well
approached by means of some of the available methods for coding, such as
Huffman's, etc.).
Then, the theoretical compression rate is defined as:
\begin{displaymath}
c_{\mbox{\ssz r, opt}}\equiv \frac{\mbox{average length per code unit}}
{\mbox{Shannon entropy per code unit}}
\end{displaymath}
Of course, for this quotient to make sense, both quantities
should be referred to the same type of code divisions
(e.g. words, data values, blocks, packets, etc.)
and must be written in the same length units (e.g. bits).

Thus, our problem entails the entropy of the stochastic
process generating the noise under consideration.
In our case, this noise will be the result of
a Gaussian proccess with a specific power spectrum.
Its outcome shall be represented by a random variable $\eta$, which
can be assumed to be stationary in wide sense. The discrete set of 
$\eta(t)$-values
for successive $t$ increases will be treated like the components
of a multidimensional Gaussian variable with the power spectrum in question.
Most of the time, we will deal with a bandwidth-limited spectrum,
i.e., one where the frequencies are limited by an upper and a lower
limit. 
Examining the associated Shannon entropy,
we shall study the hypothetic chances of compressing the sort of data
sequences generated by such processes.
In particular, we will consider Gaussian white noise,
Gaussian noise with correlation of the $1/f$-type, and
Gaussian noise with a mixed correlation of the type white-noise $+1/f$-noise.

In general,
the compression rate $c_r$ for finite sequences of symbols
that have been encoded is usually defined as the quotient between the
sequence lengths before and after the encoding process
---${L}_{\mbox{\ssz i}}$ and ${L}_{\mbox{\ssz f}}$, respectively--- i.e.,
$\ds c_r= { {L}_{\mbox{\ssz i}} \over {L}_{\mbox{\ssz f}} }$.
If $\{ a_j \}$ and $\{ \alpha_j \}$ ($j=1, \dots, {\cal N}_s$) denote the initial
and final ---or encoded--- sets of symbols,
their average lengths are
$$
\begin{array}{lll}
\ol{L}_{\mbox{\ssz i}}&=&\ds\sum_{j=1}^{{\cal N}_s} p_j \, L( a_j )  , \\
\ol{L}_{\mbox{\ssz f}}&=&\ds\sum_{j=1}^{{\cal N}_s} p_j \, L( \alpha_j )  ,
\end{array}
$$
where 
$p_j$, $L(a_j)$ and $L(\alpha_j)$ give the probability of the $j$th
symbol and its length in bits before and after encoding, respectively.
When the sequences are long enough, the rate $c_r$
can be replaced with the quotient between the
initial and final average lengths per symbol in the way
$
\ds c_r \simeq
{ \ol{L}_{\mbox{\ssz i}} \over \ol{L}_{\mbox{\ssz f}} } .
$
We shall assume
$L(a_j)=\ol{L}_{\mbox{\ssz i}}$ $\forall j$, i.e.,
that the initial data representation consists of
symbols of the same length.

Shannon's first theorem (also called noiseless coding theorem, see e.g. \cite{Welsh,BK}) 
provides theoretical
lower (and upper) bounds to the final length per symbol in the way
$H \le \ol{L}_{\mbox{\ssz f}} \  ( \le H+1 )$, where $H$ is the Shannon entropy
\beq
H= -\sum_j p_j \log_2( p_j ) .
\label{HSh}
\eeq
An efficient coding
method will have to approach equality to the lower bound.
For one dimension,
the Huffman scheme is known to be reasonably close
\footnote{To give an idea of this closeness, let's quote a bound
found in \cite{Galla}:
calling $r\equiv \ol{L}_{\mbox{\ssz f}} -H$,
and $p_{\mbox{\ssz max}}=\mbox{max}( \{ p_j \} )$, then
$r \le p_{\mbox{\ssz max}} + \log_2\left( 2 \log_2(e) \over e \right)
= p_{\mbox{\ssz max}} + 0.086$.}
(see also the performance of other methods such as the Rice algorithm
in \cite{RiceIEEE}).
Thus, the compression ratio will satisfy
$\ds c_r \simeq
{ \ol{L}_{\mbox{\ssz i}} \over \ol{L}_{\mbox{\ssz f}} }
\le { \ol{L}_{\mbox{\ssz i}} \over H }$,
being the equality the {\it optimal} case, given by
\beq
\ds c_{\mbox{\ssz r, opt}} \equiv {  \ol{L}_{\mbox{\ssz i}} \over H  } .
\label{Cropt}
\eeq

Let's consider the case of an $N$-dimensional (vector) random variable.
Since the probabilities must be now referred to a multivariate
distribution, \req{HSh} is generalized to
\beq
H_N= -\sum_{j_1,\dots, j_N} p_{j_1,\dots, j_N} \log_2( p_{j_1, \dots, j_N} ) .
\label{HNSh}
\eeq
We shall suppose that each of its components is a
one-dimensional random variable of the same type.
In addition, 
there might exist possible correlations among these components.
There is a well-known inequality for
any $N$-dimensional random variable $\vec{\eta}=(\eta_1, \dots, \eta_N)$
(Gaussian or not)
relating the joint Shannon entropy $H_N$
and the individual Shannon entropies of each
component, $H_{1}(\eta_j)$, $j=1,\dots,N$, which reads
\beq
H_N(\eta_1, \dots, \eta_N) \le H_1(\eta_1)+\dots+H_1(\eta_N) ,
\label{HNH1}
\eeq
or, equivalently,
\beq
h(\eta_1, \dots, \eta_N) \equiv {H_N(\eta_1, \dots, \eta_N)\over N}
\le {1 \over N}\sum_{j=1}^N H_1(\eta_j) ,
\label{defh}
\eeq
where $h$ denotes the joint Shannon entropy per component.
Unlike $H_N$, $h$ does not grow extensively by merely increasing $N$.
When $\eta_1, \dots, \eta_N$ are all of them of the same type,
\req{defh} reduces to $\ds h \le H_1$.
Defining the initial length {\it per component} $\ol{l}_{\mbox{\ssz i}}$
as the analogue of $\ol{L}_{\mbox{\ssz i}}$ for each vector component,
eq. \req{Cropt} may be rewritten as
\beq
c_{\mbox{\ssz r, opt}} \equiv {  \ol{l}_{\mbox{\ssz i}} \over h  } .
\label{defcrpc}
\eeq
It is essential to note that the equality in \req{defh} is satisfied
if and only if the $N$ components $\eta_1, \dots, \eta_N$ are independent.
Therefore, for independent variables of the same type,
$h= H_{N=1}$, and it is enough to study the $N=1$ case.

Observe that for a uniform distribution, where $p_j= 1/{{\cal N}_s}$, we have that
$h=\ol{l}_{\mbox{\ssz i}}=\log_2({{\cal N}_s})$;
so, no compression is possible ($c_{\mbox{r, opt}}=1$).

\section{One-dimensional Gaussian variable}

We will try to find this theoretical rate for a
zero-mean Gaussian white noise $\eta$,
---whose probability density will be called $f(\eta)$---
with variance equal to $\sigma$,
and whose values are discretized or `quantized' to a given resolution.
When discretizing,
we gather results into intervals of some fixed width, which
shall be denoted by $\Delta\eta$.
If this width is small enough, we may assume that all the values
that have fallen into the same interval have, roughly, the same
probability. Thus, to each interval we assign a `probability' value
as follows
$$
\mbox{$\eta$ in the interval around $\chi$ } 
\left( \chi-{\Delta\eta \over 2}, \chi+{\Delta\eta \over 2} \right)
\longrightarrow
\left\{ \begin{array}{lll}
p(\chi) \, \Delta\eta &=&\ds\int_{\ds\chi -{\Delta\eta\over 2}}^{\ds\chi +{\Delta\eta\over 2}}
d\zeta \, f(\zeta) \\
&\simeq &\ds f(\chi) \, \Delta\eta =
{ e^{\ds -{\chi^2 \over 2 \sigma^2}} \over \sqrt{2 \pi \sigma^2} } \, \Delta\eta
\end{array} \right.
$$
This will be done for each 
$\eta^{(j)}$, with $\eta^{(j)}= j \Delta\eta$, $j \in {\bf Z}$.
Each interval will be called
$I^{(j)}= \left( \eta^{(j)}-{\Delta\eta \over 2}, \eta^{(j)}+{\Delta\eta \over 2} \right)$.
In order to properly talk about probabilities, the set should be
well normalized.
Therefore,
we write the probability that $\eta$ takes a value in $I^{(j)}$ as
\beq
p_j \equiv p[ \eta \in I^{(j)} ] =
{ p(\eta^{(j)}) \over {\ds \sum_n p(\eta^{(n)}) } } =
{ e^{\ds -{j^2 \, (\Delta\eta)^2 \over 2 \sigma^2}} \over Z } ,
\label{defpj}
\eeq
where
\beq
Z= \sum_{n={-\infty}}^{\infty} e^{\ds -{n^2 \, (\Delta\eta)^2 \over 2 \sigma^2 }} .
\label{defZ}
\eeq
This $Z$, introduced in order to fulfil the normalization condition,
may be also regarded as the partition function of a system
with energies $\{ E_n= \pi n^2 \}$ at
temperature $\ds T={2 \pi \sigma^2 \over (\Delta\eta)^2}$,
with adequate new units for the Boltzmann constant.

In order to calculate the ideal compression rate, we need to find the Shannon
entropy \req{HSh}. Since $N=1$, $H= H_{1}= h$, and
the result (see subsec. \ref{subsecApp1d} in the appendix) is
\beq
h=\log_2 \left[ \sqrt{ 2 \pi e} \, {\sigma \over \Delta\eta} \right]
+{\cal O}\left( 
{2 \pi \sigma^2 \over (\Delta\eta)^2} \, e^{-{2 \pi^2 \sigma^2 \over (\Delta\eta)^2}}, \dots
\right)
\label{H12l}
\eeq
which depends on $\sigma$ and $\Delta\eta$ only trhough the 
dimensionless quotient
\beq
{\Delta\eta \over \sigma}\equiv \lambda 
\label{defla2}
\eeq
Thus, the smaller $\lambda \sim 1/\sqrt{T}$ (the higher the temperature) 
the larger the entropy  $h$.
Compare this with the result of a na{\"\i}ve integration without
discretization, which would be 
$\log_2 \left( \sqrt {2 \pi e \sigma^2} \right) \equiv h_{\mbox{\ssz cont}}$.
In the $\lambda\to 0$ limit the exponentially small corrections vanish,
but the logarithm of $\lambda$ diverges. Thus,
\beq 
h \simeq h_{\mbox{\ssz cont}}- \log_2(\Delta\eta)
\label{hhcontn0}
\eeq
(see explanation in refs. \cite{Ko} or \cite{Papou}, or in app. \ref{appcrv}, or
our own comments below, after eq.\req{HHcont}).

Let's now write the initial mean length as
$\ol{l}_{\mbox{\ssz i}}=\log_2( {\cal N}_s )$.
This means that, using a suitable binary representation,
${\cal N}_s$ is the number of
{\it effectively distinct} $\eta$-values
that can be considered (although  $\ol{l}_{\mbox{\ssz i}}$ is
an integer only when ${\cal N}_s$ is an exact power of 2,
these variables will be treated as if they were real).

\begin{figure}
\centerline{\psfig{figure=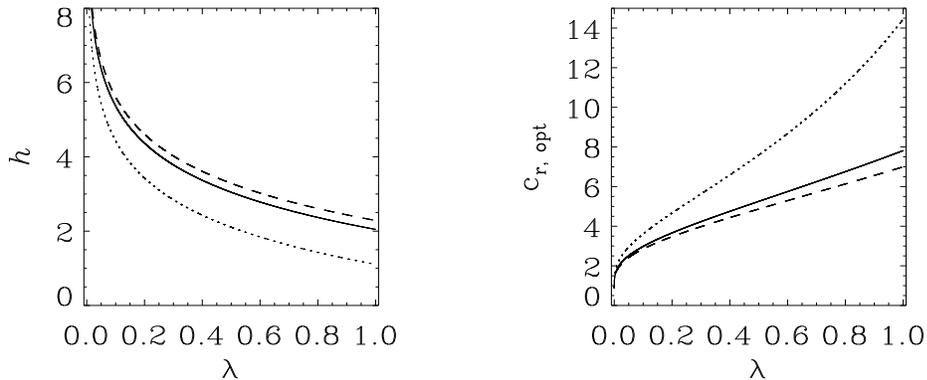,width=14cm,height=6cm}}
\caption[junk]{
Shannon entropy per component $h$ (left) and associated
optimal compression rate $c_{\mbox{\ssz r, opt}}$ (right)
---by formulas \req{defcrpc},\req{crfli}---
as functions of the discretization parameter $\lambda=\Delta\eta/\sigma$,
for a fixed $\ol{l}_{\mbox{\ssz i}}=16$ bits.
The three curves correspond to
$n_p=0$ with $P(\omega)=A=1$ sec. (solid line),
a combination of $n_p=0$ and $n_p=-1$
of the form $P(\omega)=A(A_0+\omega_0/|\omega|)$ with
$A_0=1$, $\omega_0=\omega_{\mbox{\ssz Max}}/10$ (dashed line),
and to $n_p=-1$ for $P(\omega)=A\omega_0/|\omega|$
with the same value of $\omega_0$ (dotted line).
No equal $\sigma_{1-p}$-constraint has been imposed.}
\label{fig:1a}
\end{figure}

First, we can imagine a process in which the initial length
per symbol $\ol{l}_{\mbox{\ssz i}}$ has been fixed independently of
$\Delta\eta$ (this could be the case when we are worried about 
instabilities of the signal). Then, 
the optimal compression rate would just be the quotient
\beq
c_{r, \mbox{\ssz opt}} \equiv {  \ol{l}_{\mbox{\ssz i}} \over h  }
 \simeq { \ol{l}_{\mbox{\ssz i}} \over
\log_2(\sqrt{ 2 \pi e}/\lambda)}.
\label{crfli}
\eeq
So that the larger we can make $\lambda$, without loss of relevat information,
the larger the compression. If the final sensibility ${\cal S}$ 
we need is  
obtained from some later average of $M$ measurements of this noise $\eta$, then
we can make $\lambda \simeq 1$ as far as  $M \simgt (\sigma/{\cal S})^2$. In this
extreme case the compression can be as large as $c_{r, \mbox{\ssz opt}} 
\simeq \ol{l}_{\mbox{\ssz i}}/2.047$, eg. $c_{r, \mbox{\ssz opt}}=7.8$ for 16 bits
symbols.  Fig. \ref{fig:1a} shows (as continuous lines) the entropy $h$ and the
compression $c_{r, \mbox{\ssz opt}}$ 
as a function of $\lambda$.

\bigskip

Another possibility is to work with $\ol{l}_{\mbox{\ssz i}}$ as
a function of $R$ and $\Delta\eta$.
We suppose that the values of
our random variable $\eta$ span a range 
$R\equiv\mbox{max}(\eta)-\mbox{min}(\eta)$. 
Assuming our discretization to be {\it linear}, it is clear that
\beq
R= {\cal N}_s \Delta\eta
\label{RNid}
\eeq
and, therefore,
\beq
\ol{l}_{\mbox{\ssz i}} = \log_2\left( R \over \Delta\eta \right) .
\label{LiRd}
\eeq
Formulae \req{defcrpc} and \req{H12l}-\req{LiRd}
enable us to put $c_{\mbox{\ssz r, opt}}$ as a
function of either $\Delta\eta$ or  $\ol{l}_{\mbox{\ssz i}}$.
Then,
\beq
c_{r, \mbox{\ssz opt}}  \simeq
{\log_2(R) - \log_2(\Delta\eta) \over h_{\mbox{\ssz cont}} - \log_2(\Delta\eta)}
={ \ol{l}_{\mbox{\ssz i}} \over
h_{\mbox{\ssz cont}} +  \ol{l}_{\mbox{\ssz i}} - \log_2(R) }
={ \ol{l}_{\mbox{\ssz i}} \over 
\ol{l}_{\mbox{\ssz i}} 
+ \log_2\left[ \sqrt{2 \pi e \left(\ds \sigma\over R \right)^2 } \right] }.
\label{crfnct}
\eeq
If we limit $R$ to a given number of $\sigma$'s ---say $N_0$---
around the origin, only the values in
$(-N_0\sigma, N_0\sigma)$ will be taken into consideration. Thus,
$R= 2N_0\sigma$, and we can further write
\beq
c_{r, \mbox{\ssz opt}}  \simeq
\ds { \ol{l}_{\mbox{\ssz i}} \over
\ol{l}_{\mbox{\ssz i}} + \ds \log_2\left( \sqrt{ 2 \pi e } \over 2 N_0 \right) }
={1 \over 1+
{\ds \log_2\left( \sqrt{ 2 \pi e } \over 2 N_0 \right) \over
\ds \log_2\left( 2 N_0 \sigma \over \Delta\eta \right) } } .
\eeq
Note that $c_{r, \mbox{\ssz opt}}$ cannot be larger than one
if 
$N_0 \le N_{0 \, \mbox{\ssz crit}}\equiv{\sqrt{2 \pi e} \over 2 }\simeq 
2.0664$. 
This is interpreted as a critical size of the acceptable range.
On the other hand, by taking larger and larger values of $N_0$ one could
achieve arbitrarily high compression rates, but this would
mean to collect sufficiently meaningful amounts of data very far from the mean.
This could correspond to rare events which might not follow the Gaussian 
distribution.\footnote{Note that there is no contradiction here because 
even in the presence of non-Gaussian rare events, the bulk of the data might still
be well described by a Gaussian so that our estimations could still yield
a good approximation.} 
In general, a reasonable choice would be some
$N_0$ moderately above $N_{0 \, \mbox{\ssz crit}}$, but this depends critically
on the subsequent data analysis we want to carry on with these data.

\begin{figure}
\centerline{\psfig{figure=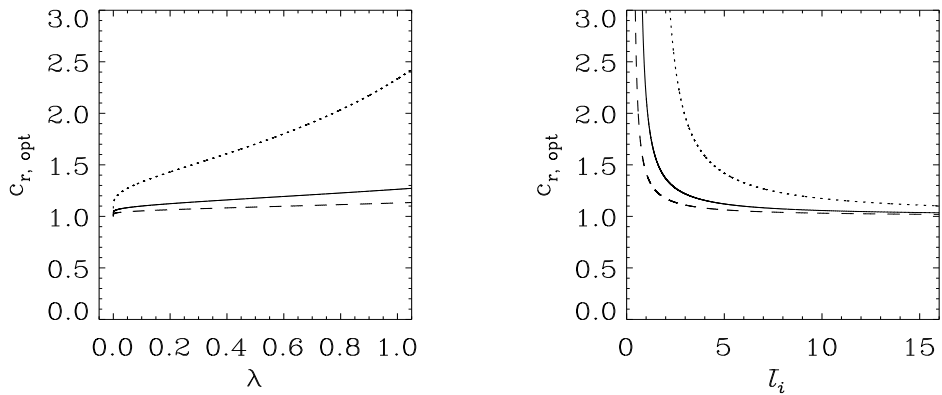,width=14cm,height=6cm}}
\caption[junk]{
Optimal compression rate
$c_{\mbox{\ssz r, opt}}$ 
---by formulas \req{LiRd},\req{crfnct}---
as functions of the discretization interval $\lambda=\Delta\eta/\sigma$ (left)
and as functions of the initial mean length in bits
$\ol{l}_{\mbox{\ssz i}}$ (right).
The three curves have the same parameters as in Fig. \ref{fig:1a} but,
this time, with variable $\ol{l}_{\mbox{\ssz i}}$ and $R=2N_0\sigma$, $N_0=3$.
Note that $c_{\mbox{\ssz r, opt}}$ has a divergence
for small $\ol{l}_{\mbox{\ssz i}}$, which comes from
the vanishing of $h$.}
\label{fig:1b}
\end{figure}

In Figure \ref{fig:1b} we show  $c_{r, \mbox{\ssz opt}}$
for the case of white noise (continuous line) with  
$R=2 N_0 \sigma$, $N_0=3$.
The main difference from Fig.\ref{fig:1a} is that in the 
former $\ol{l}_{\mbox{\ssz i}}=16$ bits
while in Fig.\ref{fig:1b} we choose $\ol{l}_{\mbox{\ssz i}}$ according to $\Delta\eta$
as in eq.\req{LiRd} with $N_0=3$.
Although the distance of three sigmas is already a long way from the mean,
the compression rates found are rather small. In Fig. \ref{fig:1b} we also show
(right panel) $c_{r, \mbox{\ssz opt}}$ as a function of 
$\ol{l}_{\mbox{\ssz i}}$, showing how the compressibility increases as
$\ol{l}_{\mbox{\ssz i}}$ gets small.

\section{Multidimensional case: Gaussian stochastic processes}

\subsection{Uncorrelated Gaussian variables: white noise}

Suppose now that we have $N$ uncorrelated Gaussian variables with different variances
$\sigma_1, \dots, \sigma_N$. 
Although we shall keep this general notation, we are only interested  
in  processes where
$\sigma_1=\dots=\sigma_N$, which is the case of a Gaussian stochastic process
stationary in wide sense. 
As long as these $N$ variables are uncorrelated,
we have to apply \req{HNH1} as an
equality, which, combined with \req{H12l} and equally quantizing in all 
dimensions, gives the joint entropy
\beq
H\equiv H_N = \log_2\left[
\sqrt{ \left( 2 \pi e \over (\Delta\eta)^2 \right)^N
\prod_{m=1}^N \sigma_m^2 } 
\right]
+{\cal O}\left( N \,
{2 \pi \sigma^2 \over (\Delta\eta)^2} \, e^{-{2 \pi^2 \sigma^2 \over (\Delta\eta)^2}}, \dots
\right) .
\label{HNuc}
\eeq
Here we may interpret
\beq
\prod_{m=1}^N \sigma_m^2 = \Det(C_0),
\label{DetC0}
\eeq
where $C_0$ is the diagonal matrix
\beq
C_0=\mbox{diag}( \sigma_1^2, \dots, \sigma_N^2 ) .
\label{C0}
\eeq
It is usefull to define an {\it effective} variance as:
\beq
\sigma_0^2 \equiv \Det^{1/N}(C_0),
\label{sigmae0}
\eeq
so that when $\sigma_1=\dots=\sigma_N$ we will have
$\sigma_0=\sigma_1=\dots=\sigma_N$, which also agrees with our
definition of the 1-point sigma $\sigma_{1p}$ below eq.\req{defsigma1p}.

Thus, the entropy per component is conveniently written as in the previous case
\req{H12l}:
\beq
h=\log_2 \left[ \sqrt{ 2 \pi e} \, {\sigma_0 \over \Delta\eta} \right]
+{\cal O}\left(
{2 \pi \sigma^2 \over (\Delta\eta)^2} \, e^{-{2 \pi^2 \sigma^2 \over (\Delta\eta)^2}}, \dots
\right) \equiv h_0(\sigma_0)
\label{defh0}
\eeq
where the 0-subscript means that this is the uncorrelated case.
As we shall see below, to deal with correlations will just mean
the replacement of $C_0$ with a new correlation matrix ---say $C$
in \req{sigmae0}.

We have done
simulations of Gaussian noise
with $\sigma_1=\dots=\sigma_N$ and the data have been represented with
a fixed $\ol{l}_{\mbox{\ssz i}}=8$ bits.
The $N$-dimensional
variable is then compressed by the Huffman and arithmetic methods, 
and the compression rate
$c_{\mbox{\ssz r}}^H$ is found as the quotient between the sizes of the
initial and the compressed files.
This actual compression rate is then compared to the optimal one, i.e., to
$c_{\mbox{\ssz r, opt}}={ \ol{l}_{\mbox{\ssz i}} \over h}$. 
The results are presented in Table 1.
The agreement is better as $N$ increases. The explanation
is that, in practice, the compressed files take up some further space
for storing the conversion tables between both symbol sets. Obviously,
since the number of different symbols is fixed
---$2^{ \ol{l}_{\mbox{\ssz i}}}=256$---
the relative contribution caused by the size of these tables decreases as $N$
grows.

\begin{table}
\begin{center}
\begin{tabular}{|r|l|r|r|r|}
\hline
$N$&$\lambda\equiv{\Delta\eta\over\sigma}$&
$c_{\mbox{\ssz r}}^{\mbox{\ssz H}}$ &
$c_{\mbox{\ssz r}}^{\mbox{\ssz A}}$&
$c_{\mbox{\ssz r, opt}}$ \\ \hline
  1000&0.05&1.09&1.09&1.26 \\ \hline
 10000&0.05&1.23&1.23&1.26 \\ \hline
100000&0.05&1.25&1.25&1.26 \\
      &0.2&1.81&1.83&1.83 \\
      &0.4&2.34&2.36&2.37 \\
      &0.6&2.81&2.85&2.85 \\
      &0.8&3.27&3.31&3.32 \\
      &1.0&3.67&3.69&3.80 \\ \hline
\end{tabular}
\caption[junk]{Comparison of optimal compression rates 
$c_{\mbox{\ssz r, opt}}$ 
with actual rates from simulated Gaussian white noise
compressed with implementations of
the Huffman ---$c_{\mbox{\ssz r}}^{\mbox{\ssz H}}$---
and arithmetic ---$c_{\mbox{\ssz r}}^{\mbox{\ssz A}}$--- methods.
These values are roughly a half of those in the solid-line curve
of Fig. \ref{fig:1a}, as $\ol{l}_{\mbox{\ssz i}}$ is now equal to 8,
instead of 16.}
\label{actualcr}
\end{center}
\end{table}

\subsection{Gaussian variables with correlation: coloured noise}

Now, suppose that
we have an $N$-dimensional variable $\vec\eta=( \eta_1, \dots, \eta_N)$
whose components are
{\it correlated} according to the entries of some covariance matrix $C$ .
By mathematical definition,
$
C( \eta_j, \eta_k )=
\langle ( \eta_j -\ol{\eta_j} ) ( \eta_k-\ol{\eta_k} )^* \rangle ,
$
where $\langle \dots \rangle$ denotes statistical average, and
$\ol{\eta_j} \equiv \langle \eta_j \rangle $.
In the case of zero-mean variables, it reduces to
\beq
C( \eta_j, \eta_k )= \langle \eta_j {\eta_k}^* \rangle \equiv C_{jk}
\label{defC}
\eeq
(for the changes to be made when the mean is not zero, see sec. \ref{Appnzm}).
In practice, a discretization or {\it shot} noise fluctuation could be added
and the theorical correlation would be changed to 
$\ds C_{jk}= \langle \eta_j {\eta_k}^* \rangle
+ {1 \over \langle N \rangle}$. In general, this is of little interest
as it just amounts to a constant increase  of the power
spectrum. The values of $\vec\eta$ can 
correspond to continuous random 
variable $\eta=\eta(t)$ sampled in $N$ time intervals
($\vec\eta=\eta(\vec t)$).
For a wide sense stationary stochastic process we have that
$C_{jk}=C_{j-k}$ can only be a function of $j-k$, i.e.,
the covariance matrix is a Toeplitz matrix.

A sequence of a Gaussian stochastic process 
has a joint probability density given by
$$
f( \eta ) \propto 
e^{\ds -{1 \over 2} \vec{\eta}^{ \  T} \, {C}^{-1} \, \vec{\eta}^* } .
$$
In the absence of correlations $C$ is just the $C_0$ of \req{C0} and
therefore
${C}^{-1}=$ diag$( 1/\sigma_1^2, \dots, 1/\sigma_N^2)$, but now
we expect the presence of nonvanishing off-diagonal coefficients.
We may assume that all the $\vec\eta$ components are real.
Each dimension will be discretized in the same way as for the
one-dimensional case. Therefore, we will consider the joint probabilities
\beq
\begin{array}{lll}
p_{j_1, \dots, j_N} &\equiv&
p[ \eta_1 \in I_{j1}, \dots , \eta_N \in I_{j_N} ] \\
&=&\ds {1 \over Z} \
e^{\ds -{(\Delta\eta)^2 \over 2}
(j_1, \dots, j_N)^T \, {C}^{-1} \, (j_1, \dots, j_N) } ,
\end{array}
\eeq
where the normalizing quantity $Z$ is given by
\beq
Z= \sum_{n_1, \dots, n_N}
e^{\ds -{(\Delta\eta)^2 \over 2}
(n_1, \dots, n_N)^T \, {C}^{-1} \, (n_1, \dots, n_N) } .
\label{ZN1}
\eeq
The ensuing Shannon entropy (see subsec. \ref{AppNd} in the appendix) is
\beq
H\equiv H_N= \log_2\left[
\sqrt{ \Det\left( {2 \pi e \over (\Delta\eta)^2} \, {C} \right) }
\right]
+{\cal O}\left( 
N \, {2\pi\sigma^2 \over (\Delta\eta)^2 } e^{-{2\pi^2\sigma^2\over(\Delta\eta)^2}} , \dots
\right) ,
\label{HlogDet}
\eeq
where this  $\sigma$ is given in section \req{AppNd}.
Note that the next-to-leading terms are, again, exponentially small, 
and their typical size can be adequately expressed as a function of a
dimensionless parameter ${\Delta\eta \over \sigma}\equiv\lambda$.
Before going on, some comments are in order. The previous relation
can be rewritten in the form
\beq
H = \log_2\left[ \sqrt{ \Det( 2 \pi e \, C ) } \right]
-N \log_2(\Delta\eta)
+\mbox{exponentially small part}
\label{seplogdelta}
\eeq
The first term on the r.h.s. is just the result
of having calculated $H$ after replacing the multiple sum in \req{ZN1}
with a multiple integral. Therefore, we shall call it $H_{\mbox{\ssz cont}}$.
Further, in the continuum limit, $\lambda\to 0$ and the exponential corrections
should vanish. This leads to
\beq
H = H_{\mbox{\ssz cont}} - N \, \log_2( \Delta\eta ) ,
\label{HHcont}
\eeq
When an entropy associated to a discretization of width $\Delta\eta$
is compared with its continuous version, we realize that we gain
$N$ times the `information' leaked by mistaking a single element 
of unit length for an interval of size $\Delta\eta$, which is
$N \, [ -\log_2(\Delta\eta)+\log_2(1)] = -N \log_2(\Delta\eta)$.
In terms of entropy per component, \req{HHcont} becomes
$h = h_{\mbox{\ssz cont}} - \log_2( \Delta\eta )$, which
generalizes \req{hhcontn0}, as now $h_{\mbox{\ssz cont}}$ has the same
expression as in \req{hhcontn0} but changing $\sigma $ by $\sigma_e$ .
Furthermore, there is a critical $\Delta\eta$-value for which the whole $h$
vanishes.
When this happens,
the discretization is so coarse that the little resolution
kept is not enough to store any effective information at all.

Another convenient way of writing the entropy per component is
\beq
h=\log_2 \left[ \sqrt{ 2 \pi e} \, {\sigma_e \over \Delta\eta} \right]
+{\cal O}\left(
{2 \pi \sigma^2 \over (\Delta\eta)^2} \, e^{-{2 \pi^2 \sigma^2 \over (\Delta\eta)^2}}, \dots
\right)
\label{defhcorr}
\eeq
were we have now that the {\it effective} variance is:
\beq
\sigma_e^2 \equiv \Det^{1/N}(C),
\label{sigmae}
\eeq
These expressions generalize to correlated variables the result
in eq. \req{defh0} for $h_0$  by just
replacing $C_0$ with $C$ and $\sigma_0$ with $\sigma_e$.
Thus, for a general covariance matrix $C$ we only need
to find $\sigma_e$ above to obtain the corresponding
entropy.

\subsubsection{Calculation of Det(C)}

The next task is the calculation of the determinant of ${C}$.
Going to Fourier space 
---see sec.\ref{secFS}, subsec.\ref{subsecDetC}--- one obtains the
relation
\beq 
\Det(C) = 
\left( { \Delta\omega \over 2 \pi \, \Delta t } \right)^N 
\Det\left( \widehat{C} \right) ,
\label{DetChC}
\eeq
where $\widehat{C}$ is the Fourier-space representation of $C$,
$\Delta t$ is the Fourier time sampling interval, $\Delta\omega$
the associated frequency interval and, taking into account that
$N$ samples are considered, $\ds\Delta\omega={2\pi \over N\Delta t}$.
In order to find concrete results, some sort of hypothesis
on $\widehat{C}$ has to be made. Here we consider
stationary (or homogeneous) processes, for which the
the covariance matrix is a Toeplitz matrix, and therefore
 $\widehat{C}$ is diagonal ---see subsec.\ref{subsecPS}--- so that
$\langle \widehat\eta(\omega) \, \widehat\eta^*(\omega') \rangle =
P(\omega) \, \delta_{\mbox{\ssz Dirac}}(\omega-\omega')$,
whose discrete version yields:
\beq
\ds \widehat{C}_{jk} = P(\omega_j) \, 
{\delta_{jk}\over{\Delta\omega}} ,
\label{CP}
\eeq
 i.e., $\widehat{C}$ is a {\it diagonal} matrix.
In all these cases, the problem
boils down to the properties of the $P(\omega)$ function.
If we denote by $P$ the diagonal matrix:
\beq
P\equiv\mbox{diag}( P(\omega_{-N/2}), \dots , P(\omega_{N/2}) ) .
\label{defMP}
\eeq
we can write the {\it effective} rms correlation 
$\sigma_e$ that appears in \req{defhcorr} by:
\beq
\sigma_e^2 \equiv \Det^{1/N}(C) = {1\over{2\pi\Delta t}}
\, \Det^{1/N}(P) =
{1\over{2\pi\Delta t}} \, \left[ \prod_j P(w_j) \right]^{1/N}.
\label{sigmae2}
\eeq
The white noise case
corresponds to the constant power spectra $P(w)=A$
and the matrix $P$ is proportional to the identity. In this case,
\beq
\sigma_e^2 = \sigma_0^2 = {A\over{2\pi\Delta t}},
\label{sigmae02}
\eeq
showing that the
larger the sampling interval $\Delta t$ the smaller
the variance, as expected.

We can also express the entropy as a difference from
the entropy $h_0$ of a white noise spectrum of amplitude $P=A$ by:
\beq
h =\ds h_0+ 
{1 \over N}\log_2\left[ \sqrt{ \Det\left( {1 \over A} P \right) } \right] .
\label{hh0}
\eeq
In general, given two power spectra $P_1$ and $P_2$ with effective
correlations $\sigma_{e1}$ and $\sigma_{e2}$, the entropy differences
are given by:
\beq
h_2-h_1 = \log_2 \left[\sigma_{e2}\over{\sigma_{e1}} \right]
= \log_2\left[ {\Det^{1/2N}(P_2) \over \Det^{1/2N}(P_1)} \right]
={1 \over N}\log_2\left[ \sqrt{ \Det\left( P_2 P_1^{-1} \right) } \right] .
\label{h2h1}
\eeq

\ni\ul{Entropy comparison for equal-$\sigma_{1p}$ processes}.
 From the expression above
\req{sigmae2} it is clear that $\sigma_e^2$ is linearly
proportional to the amplitude of the power spectrum $P(w)$,
so that $h$ will depend (logarithmically) 
on the normalization of $P(w)$.
It is interesting to compare the entropy for different shapes
of $P(w)$ which have been normalized in the same way.
Here we will consider the case where we normalize $P(w)$ so that
$\vec\eta$ has the same 1-point variance. We will see that
this is equivalent to fix the traces of the $P$ matrix
\req{defMP}.

First, using eq.\req{ChC} and the properties of the trace, we get
$\ds
\Tr(C)= \left( \Delta\omega \over 2\pi \Delta t \right)
\Tr\left(\widehat{C}\right)
$.
Combining this and \req{DetChC},
\beq
\Det(C)= \left[ \Tr(C) \over \Tr\left(\widehat{C}\right) \right]^N
\Det\left( \widehat{C} \right) .
\label{DetCChTrCCh}
\eeq
Now, bearing in mind the usual definition of the 1-point
variance: $\sigma^2$,
which reads
$\ds\sigma^2_{1p} \equiv C( \eta(t), \eta(t) )$, let's introduce
\beq
\sigma^2_{1p}(C) \equiv {1 \over N}\Tr(C) = {1\over{2\pi\Delta t}} \, 
 {1 \over N}\Tr(P) .
\label{defsigma1p}
\eeq
For the case of uncorrelated variables (white noise) with equal sigma:
$\sigma_1=\dots=\sigma_N \equiv \sigma_0$, 
we have that $\sigma_{1p}=\sigma_e=\sigma_0$
in eq.\req{sigmae2}.
In general $\sigma_{1p} \ne \sigma_e$
when there are correlations.

Using this definition, we have from \req{DetCChTrCCh}:
\beq
\sigma_e^2 \equiv \Det^{1/N}(C) = \sigma_{1p}^2 \,
{\Det^{1/N}(P) \over{\Tr(P)/N }}
\label{sigmae1p}
\eeq
Inserting this result into \req{defhcorr}, we can write:
\beq
\begin{array}{lll}
h&=&\ds h_{1p}
+{1 \over 2N}\log_2\left[{ 
\Det(P)
\over \left[ \Tr(P)/N \right]^N
} \right] ,
\end{array}
\label{hhI}
\eeq
where
\beq
 h_{1p}= h_0(\sigma_{1p}) =
\log_2\left[ \sqrt{2 \pi e} \,\, {\sigma_{1p} \over \Delta\eta } \right]
+\mbox{exponentially small part}.
\label{valhI}
\eeq
These new formulae are adequate for comparing processes with the same
value of $\sigma^2_{1p}$
and different $P$'s (i.e., different power spectra).
The 1-point entropy $h_{1p}$ denotes the entropy per component of a
white noise with a variance $\sigma^2_1=\dots=\sigma_N^2=\sigma^2_{1p}$, as in this
case $P\propto I$, causing the second
term on the r.h.s. of \req{hhI} to vanish.

For any square and positive semidefinite matrix $M$,
 the inequality
${1 \over N}\Tr(M) \ge \Det^{1/N}(M)$ holds.
 Both $C$ and $P$ satisfy these
conditions. Therefore $ \sigma_e^2 \le \sigma_{1p}^2$
and $h \le h_{1p}$. The equality is achieved when
$P\propto I$, i.e., only for the white noise itself.
In any other case,
a Gaussian process with the same $\sigma_{1p}$ 
has smaller effective variance and lower entropy 
than the corresponding white noise. This is easy to understand from \req{HNH1}
or \req{defh}.

\ni\ul{Asymptotic expressions}.
When the exact form of $\Det(P)$
is not easy to obtain, we can resort to the
following procedure.
We may assume that $P(-\omega)= P(\omega)$ and that the mode
with $\omega_0=0$ has to be removed, as often happens
(this mode is related to the correlation at $t\to\infty$ and,
if one requires that the system be ergodic, it should vanish).
Then, 
\beq
 \log_2[\Det(P)] =
\ds\sum_{j= -N/2 \atop j \neq 0}^{N/2} \log_2[ P(\omega_j) ]
= 2 \sum_{j=1}^{N/2} \log_2[ P(\omega_j) ] ,
\eeq
and an application of the Euler-Maclaurin
summation formula (see e.g. \cite{AS}), leads us to the approximation
\beq
\begin{array}{ll}
\ds \sum_{j =1}^{N/2} \log_2[ P(\omega_j) ]=
&\ds{1 \over \Delta\omega}  
\int_{\omega_{\mbox{\ssz min}}}^{\omega_{\mbox{\ssz Max}}} d\omega \, \log_2[ P(\omega) ]
+{1 \over 2} \left(
\log_2[ P( \omega_{\mbox{\ssz Max}} ) ] + \log_2[ P( \omega_{\mbox{\ssz min}} ) ]
\right) \\
&\ds + \mbox{higher order terms in $\Delta\omega$}  .
\end{array}
\label{Eulersf}
\eeq
The same method can be applied to
the calculation of $\Tr(P)$, in \req{hhI}, i.e.,
\beq
\begin{array}{ll}
\ds 2\sum_{j=1}^{N/2} P(\omega_j)=2&\ds\left[
{1 \over \Delta\omega} 
\int_{\omega_{\mbox{\ssz min}}}^{\omega_{\mbox{\ssz Max}}} d\omega \, P(\omega)
+{1 \over 2}( P(\omega_{\mbox{\ssz Max}}) + P(\omega_{\mbox{\ssz min}}) )
+\mbox{higher order terms in $\Delta\omega$} \right] .
\end{array}
\label{EulersfTr}
\eeq

\ni\ul{Filters}.
Quite often, stochastic processes go through what is called a filter.
Formally, filters can be pictured as multiplicative changes
in the power spectrum. 
Therefore, everything happens as if we
had a new power spectrum function, say $P'$, coming from
the replacement
$$
P(\omega) \longrightarrow P'(\omega)=P(\omega) \phi(\omega) ,
$$
where the $\phi$ function is the frequency response of the filter itself.
Let $h'$ denote the new entropy per component.
It is immediate that the change caused by the
introduction of $\phi$ will be given by
\beq
\begin{array}{lll}
h'&=&h+h_{\phi}, \\
h_{\phi}&=&\ds{1 \over N}\log_2[\sqrt{\Det(\Phi)}], \hspace*{1cm}
\Phi=\mbox{diag}( \phi(\omega_{-N/2}), \dots , \phi(\omega_{N/2})),
\end{array}
\eeq
where $h$ denotes the entropy per component for the same process
when no filter is present.

\subsubsection{Simple power-law power spectrum}

\begin{figure}[htbp]
\begin{center}
\begin{tabular}{c}
\psfig{figure=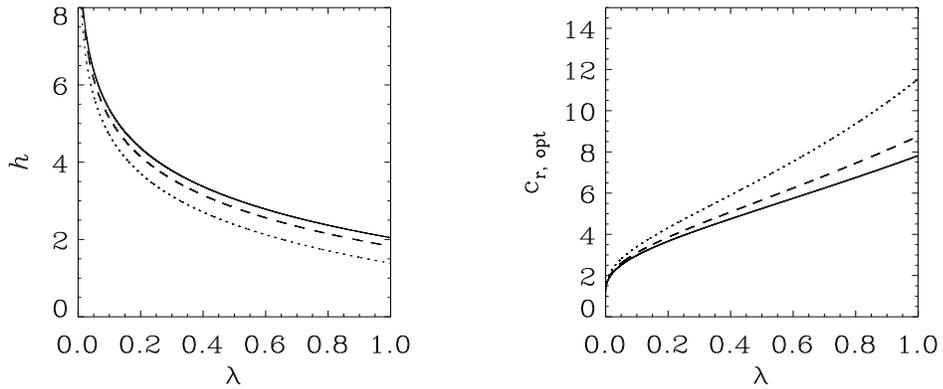,width=14cm,height=6cm}
\end{tabular} \\
\end{center}
\caption[junk]{Entropy and optimal compression rate 
for different power spectra with the same
$\sigma_{\mbox{\ssz 1p}}$, but (unlike in Fig \ref{fig:1b}) keeping
$\ol{l}_{\mbox{\ssz i}}=16$ bits fixed and $\omega_0=\omega_{min}$.
The present set of cases is:
$n_p=0$ (solid line), $n_p=-1$ (dashed line) and $n_p=+1$ (dotted line).
}
\label{fig:1c}
\end{figure}

Here, we will consider a power spectrum of the type
\beq P(\omega) = A \left({| \omega |\over{w_0}}\right)^{n_p}, 
\eeq
where $A$ is a constant that sets the overall amplitude and
$w_0$ some characteristic scale that sets the time units.
Taking into account the discrete $\omega$-values \req{omegaj} we evaluate
\beq
\Det\left( {1 \over A} P \right)=
\prod_{j} P(\omega_j)= \left( \Delta\omega \over \omega_0 \right)^{N \, n_p}
\left[ \left( N \over 2 \right) ! \right]^{2n_p}
\eeq
(where the zero mode $j=0$ has been omitted).
Making use of Stirling's approximation for large $N/2$, and using
the frequency relations \req{kMkm}, we find:
\beq
\sigma_e^2 \equiv \Det^{1/N}(C) \simeq {A\over{2\pi\Delta t}} \,
\left({\pi \over{e \, w_0 \Delta t}}\right)^{n_p}
=  \sigma_0^2 \,
\left({\omega_{\mbox{\ssz Max}}\over{e \, w_0}}\right)^{n_p},
\label{sigmaenp}
\eeq 
where $\sigma_0$ corresponds to the white noise case ($n_p=0$).
If we normalize the spectrum
at $ w_0=\omega_{\mbox{\ssz Max}}$ then
for $n_p<0$ we have that $h > h_0$ and
the optimal compression rate has  to decrease, while 
for  $n_p>0$ we have $h< h_0$. 
Some special values are given in table 2, 
and are also illustrated by Fig. \ref{fig:1a}.
However, this comparison depends on the normalization and involves noises
with different values of
$\sigma_{1p}$, as we have only changed the value of $n_p$ without
doing anything to maintain the initial $\sigma_{1p}$. In this case,
by eq. \req{defsigma1p},
\beq
\sigma^2_{1p}= {1\over{2\pi\Delta t}} \, 
{1 \over N} \,
\sum_j P(\omega_j)= {A\over{\pi N \Delta t}} \, 
\left( \Delta\omega \over w_0 \right)^{n_p}
S_{n_p}\left( N \over 2 \right) , \hspace*{1cm}
S_{n_p}\left( N \over 2 \right)\equiv \sum_{j=1}^{N/2} j^{n_p} .
\label{TrPSnp}
\eeq
Making use of \req{hhI}, we are led to
\beq
\begin{array}{lll}
h&=&\ds h_{\mbox{\ssz 1p}}
+{n_p\over N}\log_2\left[ \left( N \over 2 \right)! \right]
-{1 \over 2}\log_2\left[ {2 \over N} S_{n_p}\left( N \over 2 \right) \right] \\
&=&\ds h_{\mbox{\ssz 1p}}
+{n_p+1 \over 2}\log_2\left( N \over 2 \right) -{n_p\over 2}\log_2(e)
-{1\over 2}\log_2\left[ S_{n_p}\left( N \over 2 \right) \right]
+{\cal O}\left( { \log_2(N) \over N }, \dots \right) ,
\end{array}
\eeq
where the Stirling approximation has been applied.
When $n_p > -1$, we apply the Euler-Maclaurin summation formula \req{EulersfTr}
and obtain
\beq
\sigma^2_{1p}= {1\over{2\pi\Delta t}}  {A \over n_p + 1}
\ds\left( \pi \over w_0 \Delta t \right)^{n_p}
\left[ 1 + {\cal O}\left(  1 \over N  \right) \right] , \hspace*{1cm}
\ \mbox{for $n_p >  -1$}.
\eeq
For the $n_p= -1$ case may be more straightforwardly estimated by using
\beq
S_{-1}\left( N \over 2 \right)=
\Psi\left( {N \over 2} +1 \right) + \gamma
=\ln\left( N \over 2 \right) + \gamma + {\cal O}\left( 1 \over N \right ),
\eeq
where $\gamma$ is Euler's constant: $\gamma \simeq 0.57721 \dots$ .
So, $\sigma^2_{1p}$ becomes
\beq
\sigma^2_{1p}= {(A w_0) \over{2\pi^2}} \, 
\left[\ln\left( N \over 2 \right) + \gamma + {\cal O}\left( 1 \over N \right)\right]
\hspace*{1cm}\ \mbox{for $n_p = -1$}
\eeq
Then, by the previous formulas and by \req{hhI},
\beq
h= \left\{
\begin{array}{ll}
\ds h_{1p}- {n_p \over 2}\log_2(e)+{1 \over 2}\log_2(n_p+1)
+{\cal O}\left( {\log_2(N) \over N}, \dots \right),
&\mbox{for $n_p>-1$,} \\
\ds h_{1p}+ {1 \over 2}\log_2(e)
-{1\over 2}\log_2\left[ \ln\left( N \over 2 \right) + \gamma \right]
+{\cal O}\left( {\log_2(N) \over N}, \dots \right),&\mbox{for $n_p=-1$,}
\end{array}
\right.
\label{hhInpm1}
\eeq
where $h_{1p}$, given by \req{valhI},
is the entropy per component of a white noise 
with the $\sigma_0=\sigma_{1p}$. 
Note that, although it seems that $h$ diverges with $N$ for $n_p=-1$, this
is an artifact of this type of comparison with a fixed $\sigma_{1p}$. 
Although $\sigma^2_{1p}$ diverges logarithmically with $N$,
the information content does not, as $\sigma_e^2$ in eq.\req{sigmaenp} is
finite:
\beq
\sigma^2_{e}= {(A w_0) \over{2\pi^2}} \, e.
\hspace*{1cm}\ \mbox{for $n_p = -1$}
\label{sigmaenp-1}
\eeq
Some examples are illustrated by the 5th column of Table 2 
and Fig. \ref{fig:1c}.

\begin{figure}[htbp]
\begin{center}
\begin{tabular}{c}
\psfig{figure=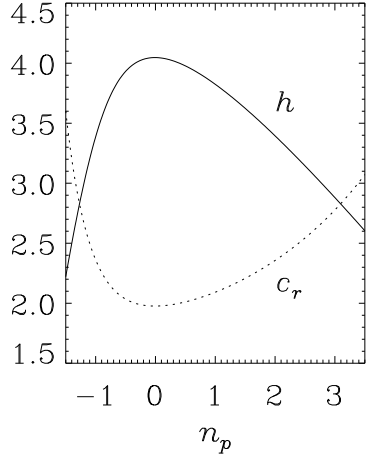,width=7cm,height=7cm}
\end{tabular} \\
\end{center}
\caption[junk]{Entropy $h$ (continuous line) and optimal compression 
$c_{\mbox{\ssz r, opt}}$ (dashed line) for $\ol{l}_{\mbox{\ssz i}}=8$ bits
as a function of the spectral index  $n_p$ (
for a power law $P(\omega) \propto \omega^{n_p}$)
with a fixed one-point variance $\sigma_{1p}$
and $\lambda_1\equiv \Delta\eta/\sigma_{1p} = 0.25$}
\label{fig:np}
\end{figure}

Fig. \ref{fig:np}, shows the entropy $h$ as a function of the spectral index
$n_p$ given by the above formulas. 
As can be seen, $h$ has a maximum at $n_p=0$,
as expected.

\subsubsection{`$f^0+1/f$' spectrum \label{sstp1of} }
\label{f0f-1}
In practice, realistic power spectra include often combinations of
several powers.
This new example corresponds to a power spectrum including two
terms: one with $n_p=0$ (white noise)
and another with $n_p=-1$ (usually called $1/f$ noise), 
which we write as
\beq
P(\omega) \equiv A \left( 1 + {\omega_{k} \over |\omega|} \right)=
A \left( 1 + {f_{k} \over |f|}\right),
\label{SkABk}
\eeq
where $f$ stands for frequency $w \equiv  2\pi f$, and $f_k$ for the
so called {\it knee} frequency, where both contributions are equal.
We shall assume that $w$ has been discretized as in the previous cases.
Because a direct evaluation of $\Det(P)$ would not be so easy now,
we shall apply the above commented approximation based on the 
Euler-Maclaurin summation formula.
After performing the integration \req{Eulersf} for the $P(\omega)$
of eq. \req{SkABk} one gets
\beq
\sigma^2_{e}= {A\over{2\pi\Delta t}} \, \left(1+{\omega_{k}\over{\omega_{\mbox{\ssz Max}}}}\right) \left[{\omega_{\mbox{\ssz Max}} + \omega_k\over{\omega_{\mbox{\ssz min}}
+\omega_k}}\right]^{\omega_k/\omega_{\mbox{\ssz Max}}}
\eeq
The correspomding entropy is just given by eq.\req{defhcorr}.
When $\omega_k << \omega_{\mbox{\ssz Max}}$ we recover the white noise
case eq.\req{sigmae02}, while in the case $\omega_k >> \omega_{\mbox{\ssz Max}}$
the $1/f$ noise dominates and we recover eq.\req{sigmaenp-1}, as expected.
We observe that a combined power spectrum \req{SkABk} with 
reasonably small $A$ is effectively equivalent to one of
the type 
$\ds P(\omega)= A\left( |\omega| \over \omega_0 \right)^{n_p}$
with an intermediate $n_p$ between 0 and $-1$. An illustration of the values of $h$
and optimal compresion for this case is shown in Fig. \ref{fig:f} and also
in Fig. \ref{fig:1a} as dashed line.

\begin{figure}[htbp]
\begin{center}
\begin{tabular}{c}
\psfig{figure=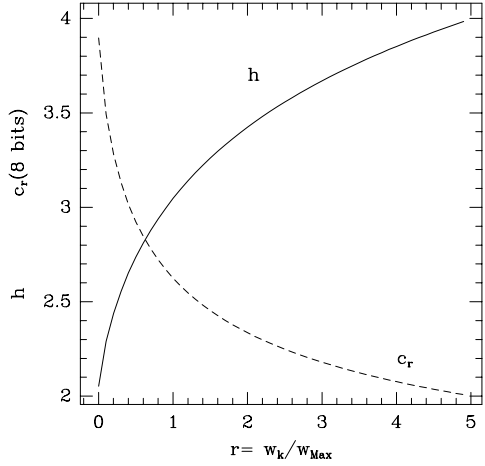,width=7cm,height=7cm}
\end{tabular} \\
\end{center}
\caption[junk]{Entropy $h$ and optimal compression 
$c_{\mbox{\ssz r, opt}}$ ($\ol{l}_{\mbox{\ssz i}}=8$ bits)
as a function of  $r \equiv \omega_k/\omega_{\mbox{\ssz Max}}$
for a `$f^0+1/f$' noise. We have chosen $\lambda=\Delta\eta/\sigma_0=1$
and symbols of $\ol{l}_{\mbox{\ssz i}}=8$ bits.}
\label{fig:r}
\end{figure}

\begin{figure}[htbp]
\begin{center}
\begin{tabular}{c}
\psfig{figure=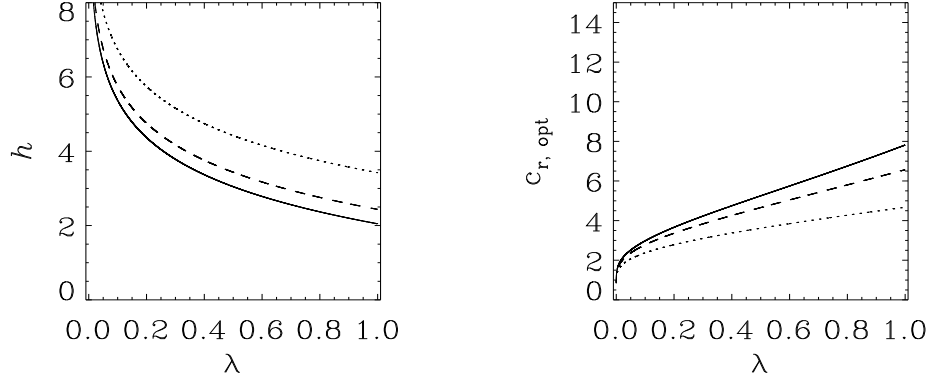,width=14cm,height=6cm}
\end{tabular} \\
\end{center}
\caption[junk]{Comparison between purely white noise (solid line)
and two processes of the type $P(f) \propto 1+f_{\mbox{\ssz k}}/|f|$
with $f_{\mbox{\ssz k}}=$10 Hz (dashed line)
and 100 Hz (dotted line), for $\ol{l}_{\mbox{\ssz i}}=$ 16 bits, and without
imposing the equal-$\sigma_{\mbox{\ssz 1p}}$ constraint. In both cases $h > h_0$,
while in the analogous example of Fig.\ref{fig:1c}
it happenned just the opposite.
}
\label{fig:f}
\end{figure}

Typically we will have that 
$\omega_{\mbox{\ssz min}} << \omega_{\mbox{\ssz Max}}$
and also  $ \omega_{\mbox{\ssz min}} << \omega_k$. In this case the only relevant 
parameter is $r \equiv \omega_k/\omega_{\mbox{\ssz Max}}$:
\beq
\sigma^2_{e}= 
{A\over{2\pi\Delta t}} \, {\left(1+ r \right)^{1+r}\over{r^r}} =
\sigma_0^2 \, {(1+r)^{1+r}\over{r^r}},
\eeq
where $r=0$ reproduces the white noise case and large $r$
reproduces the $1/f$ case ($n_p=-1$) with arbitrarily large normalization.
For $r=1$ we have that the {\it effective} variance
of the signal is four times as large as the white noise part
$\sigma^2_{e}= 4 \sigma_0^2$, so that the entropy will be one unit
larger with the combined spectrum than with the white noise
alone. Other values for $h$ and $c_r$ 
as a function of $r$ are shown in Fig.\ref{fig:r}. 
In this case $\lambda=\Delta\eta/\sigma_0=1$ so that $h_0 \simeq 2.047$
and $c_{\mbox{\ssz r, opt}} \simeq 3.91$ 
($\ol{l}_{\mbox{\ssz i}}=8$ bits) which agrees with
the values at $r=0$.

Another way to compare the two cases is to use an
equal $\sigma_{1p}$ comparison with a white noise.
In this case:
\beq
\sigma^2_{1p}\equiv {1 \over N}\Tr(C) = {A\over{2\pi\Delta t}}
\, \left[ 1 + 
{ \omega_k \over \omega_{\mbox{\ssz Max}} } S_{-1}\left( N \over 2 \right)
\right] ,
\eeq
and, using \req{hhI},
\beq
h=h_{1p} -\log_2\left[ \sqrt{ 
\omega_{\mbox{\ssz Max}} + \omega_k S_{-1}(N/2) \over
\omega_{\mbox{\ssz Max}} + \omega_k
} \right]
+{\omega_k \over \omega_{\mbox{\ssz Max}}} \, 
\log_2\left[ \sqrt{ \omega_{\mbox{\ssz Max}}+\omega_k \over \omega_{\mbox{\ssz min}}
+\omega_k } \right]
+{\cal O}\left( \log_2(N) \over N \right) ,
\label{hhIPAB}
\eeq
where $h_{1p}$ stands for the entropy per component of a Gaussian white noise
with the same 1-point variance $\sigma_{1p}^2$.
An example of this type of noise is shown in the 4th column of Table 2 and
Fig.\ref{fig:1b}.
Of course, the ${\omega_{\mbox{\ssz Max}} \over \omega_k}\longrightarrow 0$ limit
of this expression yields the $n_p=-1$ case of \req{hhInpm1} 
(see also the 5th column of Table 2).

\begin{center}
\begin{tabular}{|c|c|c|c|c|}
\hline
$\lambda=\Delta\eta/\sigma_0$&$h_0=h_{1p}$&\multicolumn{3}{|c|}{$h$} \\ \hline
& $n_p= 0$ & \multicolumn{2}{|c|}{ $f^0+1/f$} &  $n_p=-1$ \\  \hline
&$\sigma_0$  & $\sigma_0$ & $\sigma_{1p}=\sigma_0$ & $\sigma_{1p}=\sigma_0$ \\
\hline
0.05 & 6.37 & 7.37 & 5.89 & 5.71   \\
0.25 & 4.05 & 5.05 & 3.57 & 3.39    \\
0.50 & 3.05 & 4.05 & 2.57 & 2.39    \\
1.00 & 2.05 & 3.05 & 1.57 & 1.39    \\
\hline
\end{tabular}
\end{center}
\pseudocapt{
{\bf Table 2.} Shannon entropy per component $h$
for large $N$, and
several values of $\lambda =\Delta\eta/\sigma_0$.
The purely white-noise case $h_0$ for a given $\sigma_0$
and $\lambda$ are listed in column 2.
Columns 3 and 4 gives the results for a combination 
$P(\omega)= A(1+\omega_k/|\omega|)$, with 
$\omega_k=\omega_{\mbox{\ssz Max}}$ ($r=1$)
when the white noise part is
fixed to the same $\sigma_0$ (column 3) 
and when the 1-point sigma
is fixed to $\sigma_{1p}=\sigma_0$ (column 4).
In column 5 we have listed the values 
for a correlation of the $n_p= -1$ type 
$P(\omega)= A (w_0/|\omega|)$ and $\sigma_{1p}=\sigma_0$.
In the last two  cases $N=1000$.
}

We can see there how $h <h_{1p}$ when we compare spectra normalized to have
the same $\sigma_{1p}^2$, while $h >h_{1p}$ when we just add a term (1/f) to 
the (constant) white noise power spectrum. The interpretation is simple,
as shown in eq.\req{defhcorr}
the entropy is given by the {\it effective} correlation.
On the one hand, adding power always increases $\sigma_e$ (see eq.\req{sigmae2}),
and therefore $h$. But, on the other hand, $\sigma_e^2 \le \sigma_{1p}^2$
so that, when $\sigma_{1p}$ is fixed, any power spectrum gives smaller
$h$ than the white noise and, as we said above, this can be easily understood in the light of inequality \req{HNSh} . This change of behaviour can be seen 
comparing Figs.
\ref{fig:1a} and \ref{fig:1b} with Fig.\ref{fig:f}.

\subsubsection{Examples of piecewise-mixed spectra}
\label{sec:piece}
\ni{\bf 1}. Here we study the piecewise-defined spectrum:
\beq
P(\omega)=\left\{  \begin{array}{ll}
A, & \mbox{for $\omega \le \omega_L$,} \\
\ds A \, {\omega_L \over \omega}, 
& \mbox{for $\omega_L < \omega \le \omega_H$,} \\
\ds A \, {\omega_L\omega_H \over \omega^2}, 
& \mbox{for $\omega_H < \omega \le \omega_{\mbox{\ssz Max}}.$}
\end{array}
\right.
\eeq
The result of applying \req{hh0} and making asymptotic approximations
for large values of ${\omega_L\over \Delta\omega}$,
${\omega_H\over\Delta\omega}$,
and ${\omega_{\mbox{\ssz Max}}\over\Delta\omega}$ is
\beq
\begin{array}{ll}
h=&\ds h_0
+\left( 1-{ \omega_L+\omega_H \over 2 \omega_{\mbox{\ssz Max}} } \right)
\log_2(e)
-{1 \over 2}
\log_2\left( \omega_{\mbox{\ssz Max}}^2 \over \omega_L \omega_H \right) 
+\mbox{higher order terms.}
\end{array}
\label{pw3}
\eeq

\ni{\bf 2.} Another case which can be of interest is:
\beq
P(\omega)=\left\{  \begin{array}{ll}
A', & \mbox{for $\omega \le \omega_{L}$,} \\
\ds A+{B\over|\omega|} ,
& \mbox{for $\omega_{L} < \omega \le \omega_{\mbox{\ssz Max}}.$}
\end{array}
\right.
\eeq
Taking now as reference the case in which $B=0$ and $A'=A$, we may
write
\beq
\begin{array}{ll}
\ds h=h(A'=A, B=0)+&\ds
\log_2\left[ \sqrt{ 1+{B\over A\omega_{\mbox{\ssz Max}}} } \right]
-{\omega_L\over \omega_{\mbox{\ssz Max}}}
\log_2\left[ \sqrt{ 1+{B\over A\omega_{L}} } \right] \\
&\ds +{B \over \omega_{\mbox{\ssz Max}} A }
\log_2\left[ \sqrt{ 
A\omega_{\mbox{\ssz Max}}+B \over A\omega_{L}+B 
} \right]
+{\omega_L\over \omega_{\mbox{\ssz Max}}}
\log_2\left[ \sqrt{ A' \over A } \right] \\
&\ds +\mbox{higher order terms.}
\end{array}
\eeq

\section{Conclusions}

We have studied the Shannon entropy $h$ of a Gaussian discrete noise $\eta_i$
characterized by its power spectrum $P$. It amounts to
$h \simeq \log_2\left(\sqrt{2\pi e} \, \sigma_e/\Delta\eta \right)$,
where $\sigma_e= \sigma_e(P)$ is given by eq.\req{sigmae2}
and  $\Delta\eta$ is the discretization width.
The finite-$N$ corrections to this formula are exponentially small
(eqs.\req{H1app} and \req{HlogDetapp} in Appendix \ref{AppDC}).
The first thing to notice is that $\sigma_e$ changes linearly with
the amplitude of $P$, so that 
the entropy increases logarithmicaly with $P$.
For a given normalization, how does the entropy depend on the
shape of the power spectrum?
We can compare the entropy of two types of noise using
the entropy difference $\Delta h= h-h_0$. In cases with power-law spectra
$P(\omega) \propto  \left( |\omega| \over \omega_0 \right)^{n_p}$,
$\Delta h$ can be quite sensitive to the choice of $\omega_0$, whose
variations may even cause a reversal of the sign of $\Delta h$.
This type of change is due to the already commented logarithmic dependence
of $h$ on the amplitude of $P$.
If we fix the (1-point) variance of the noise, we have seen that
the maximum entropy (minimum compression) 
is the one given by white noise (or constant $P$), as expected.
For $P(\omega) \propto \omega^{n_p}$ spectra
with fixed one-point variance,
we have that the larger $|n_p|$ the smaller the entropy 
for $n_p>-1$
 (eg eq. \req{hhInpm1} and Fig.2).
Notice that when
$\Delta\eta > \sqrt{2\pi e} \, \sigma_e$ we have $h<0$ indicating that
the data have been discretized with such a low resolution
 that there is no information left.

We have defined the optimal compression rate as the ratio of the
initial average length per code unit $\ol{l}_{\mbox{\ssz i}}$
over the Shannon entropy $h$ per component:
$c_{\mbox{\ssz r, opt}} \equiv {  \ol{l}_{\mbox{\ssz i}} \over h  }$.
For a linearly discretized data set with
$\ol{l}_{\mbox{\ssz i}}=N_{\mbox{\ssz bits}}=\log_2({\cal N}_s)$ bits the optimal
compression
rate depends on the discretization width $\Delta\eta$
through a simple relation:
\beq
\ds c_{\mbox{\ssz r, opt}}\equiv {  \ol{l}_{\mbox{\ssz i}} \over h  } 
\simeq {N_{\mbox{\ssz bits}}\over{
\log_2\left(\sqrt{2\pi e} \, \sigma_e/\Delta\eta \right)}},
\eeq
The choice of
$\Delta\eta$ is in principle arbitrary and depends on what we want to do 
in the data processing of the signal (noise). 
The final compression
factors  will depend only on the ratio of these two
quantities $\lambda \equiv {\Delta\eta \over \sigma}$ and the number
of bits $N_{\mbox{\ssz bits}}$ chosen to represent the data.
Another way of writing this results is:
$\ds c_{\mbox{\ssz r, opt}}\simeq
{ \log_2(R) - \log_2(\Delta\eta) \over h_{\mbox{\ssz cont}} - \log_2(\Delta\eta) }$,
where  $R$ is the range of the random variable
and $h_{\mbox{\ssz cont}}$ is a constant depending on the type of process, which
may be interpreted as the Shannon entropy per component in the
continuum limit. In mathematical terms, $h_{\mbox{\ssz cont}}$
involves the determinant of the correlation matrix.
If the initial length $\ol{l}_{\mbox{\ssz i}}$ is held fixed,
independently of $\Delta\eta$,
the relation is just
$\ds c_{\mbox{\ssz r, opt}}(\Delta\eta) \simeq
{\ol{l}_{\mbox{\ssz i}} \over h_{\mbox{\ssz cont}} - \log_2(\Delta\eta) }$.

The purely white noise case ($n_p=0$) offers rather slight
hopes, for moderate ranges $R$. If we choose  
$R=(-N_0\sigma, N_0\sigma)$ with $N_0=3$,
and  $\lambda=\Delta\eta/\sigma= 0.25$ the compression
rate is of $c_{\mbox{\ssz r, opt}} = 1.13$
---only marginally above one--- and,
yet, this happens at the expense of losing resolution
to the extent that only four distinct values are observed within
each interval of width $\sigma$.
Less resolution than that may be too little for many applications.
One could wonder what
happens, in the opposite case, when resolution is kept at any cost.
For a binning of $2^8$ distinct intervals within the same range,
$\lambda$ has to take on such a value that
the compression rate is a meagre 1.07. Such a thinly spaced
binning means that the white noise is seen very much like a uniformly
distributed one, and has a similar uncompressibility.

On the other hand, for fixed $\sigma^2_{\mbox{\ssz 1p}}$
a negative spectral index lowers the effective information and
helps compression. Moreover, the optimal compression rate increases as
the sampling time interval decreases.
As we see in Fig. \ref{fig:1b}, when $\Delta\eta= 0.25$
the compression rate for $n_p=-1$ with the same $\sigma_{1p}$ as for the white noise
 is $\sim 1.4$.
Moreover, the difference between $n_p=-1$ and $n_p=0$
increases as the discretization parameter $\lambda=\Delta\eta/\sigma$ grows. 
However, one cannot think of
arbitrarily raising its value, as such a thing would imply
a widening of the discretization error, and an even greater
loss in resolution for the values of our variables. 

A combination of both types has also been studied by taking
a `mixed' power spectrum with $n_p=0$ plus $1/f$ (i.e. $n_p=-1$)
terms. If the coefficient of the  $n_p=0$ part is low enough,
the behaviour shown is intermediate between purely $n_p=0$ 
and purely $n_p=-1$,
and can be interpreted as if it just had an effective $n_p$ 
between both values.
When $P(\omega) \propto \left( A_0+{\omega_0\over |\omega|} \right)$,
if $A_0$ is set to 1, $h$ is not too sensitive to increases in $\omega_0$
much above the knee frequency.
On the contrary, if $\omega_0$ is kept constant, variations in $A_0$ may
easily change the sign of $\Delta h$.
As a common feature to all possible situations, one observes an
increase in compressibility as the measured data involve more
and more correlation, i.e. larger dominance of their spectral 
$f^{n_p}$-parts with $n_p \ne 0$ (see Fig. \ref{fig:np}). 

Imagine a situation of a data set that  consists of a slowly varying signal
(to be stored in $\ol{l}_{\mbox{\ssz i}}$ bits) plus large amplitude noise that dominates over
the signal on large frequencies. The signal is to be recovered 
by averaging the noise after transmission (and therefore compression)
and a careful calibration of instabilities in the noise. This is 
a commom situation for scientific measurements on-board satellites
collecting data with low signal-to-noise ratio.
In this case
the noise component can be kept with a low resolution 
and one can choose $\Delta\eta \simeq \sigma_e$ 
which gives $h \simeq 2.05$ indicating that all information 
is contained effectively in two bits. Then, high
compression rates $c_{\mbox{\ssz r, opt}} \simeq \ol{l}_{\mbox{\ssz i}}/2$
could be obtained: e.g. 
$c_{\mbox{\ssz r, opt}} \simeq 8$ for $\ol{l}_{\mbox{\ssz i}} \simeq 16$ bits.
To achieve such a high compression values in practice,
an efficient coding  method has to be used.
For one dimension, the Huffman and arithmetic schemes are known to be reasonably close
to the optimal value. When data (symbols) are correlated in a manifest way,
as the general case considered
here, other methods have to be used in combination. One of the simplest methods
that take into account correlations is run-length encoding, where the signal is
converted to a stream of integers that indicate how many consecutive symbols are 
equal (see \cite{Hunter}). This would be quite efficient in the situation
we have just mentioned.

The data discretization or `quantization' process causes a
distortion error. This issue has not been considered in the present paper,
as we have kept it outside of the scope of this study 
(i.e., we have started from a data set already quantized in a given way).
Nevertheless, the results in ref. \cite{CT} (Chap. 13)
for a univariate Gaussian source indicate that
the `best expected' average error for a
representation of a given length $\ol{l}_{\mbox{\ssz i}}$ decreases
as  $\ol{l}_{\mbox{\ssz i}}$ increases.
This confirms the intuitive idea that a
random variable like $\eta$ is better described as 
$\ol{l}_{\mbox{\ssz i}}$ grows. However, when this happens the
entropy grows too, and the compression chances are reduced.

\appendix

\section{Appendix: discrete calculations \label{AppDC}}
\subsection{One-dimensional case \label{subsecApp1d}}

First, we rewrite the $Z$ of \req{defZ} as
\beq
Z= \sum_{n={-\infty}}^{\infty} e^{\ds -{n^2 \lambda^2 \over 2}}
= \theta\left( {\lambda^2 \over 2 \pi } ; 0 \right) ,
\label{defZapp}
\eeq
where
\beq
\lambda\equiv {\Delta\eta \over\sigma }
\label{deflaapp}
\eeq
is the size of the discretization interval in units
of $\sigma$, and
$$
\theta(\beta ; m) \equiv \sum_{n=-\infty}^{\infty} n^{2m} e^{-\pi \beta n^2}
$$
is a notation for the sort of Jacobi elliptic theta functions
appearing in this calculation.

Note that the discretization has enabled us to deal with a
discrete probability set ---\req{defpj}----
thus avoiding the well-known difficulties
associated with $H$ for continuous probability distributions. 
In our own case (calling $H\equiv H_1$ all through this subsection),
$$
H= -{1 \over \ln(2)} \left\{
-{\lambda^2 \over 2} \
{ \ds \theta\left( {\lambda^2 \over 2 \pi } ; 1 \right) \over
\ds \theta\left( {\lambda^2 \over 2 \pi } ; 0 \right) }
-\ln\left[ \theta\left( {\lambda^2 \over 2 \pi } ; 0 \right) \right]
\right\} .
$$
For $m=1$, we just observe that
$$
\theta(\beta; 1) = \ds -{1 \over \pi}{d \over d\beta} \theta(\beta;0) .
$$
Using this, we arrive at
\beq
H= -{1 \over \ln(2)} \left\{
\beta \, {d \over d\beta} \ln\left[ \theta\left( \beta ; 0 \right) \right]
-\ln\left[ \theta\left( \beta ; 0 \right) \right] \right\} ,
\hspace{1cm}\mbox{with $\ds\beta = {\lambda^2 \over 2\pi }={1 \over T}$.}
\label{Hlnth}
\eeq
By \req{defZapp}, this can also be written as
\beq
H= {1 \over \ln(2)}{d \over dT} \left[ T \, \ln(Z) \right].
\label{HlnZ}
\eeq
Up to the trivial change of units ---or, equivalently, a conventional
modification of the Boltzmann constant---
$H$ is the
thermodynamical entropy $S$ of a one-particle system at temperature $T$
with partition function $Z$. In the situation we are studying,
this $Z$ is $Z(T)=\ds\theta\left( {1 \over T}; 0 \right)$
as given by eq. \req{defZ}. However, the validity of
eq. \req{HlnZ} is quite general: in fact,
for {\it any} system  with
probabilities of the form 
$$\ds p_J={ e^{-{E_J/T}} \over Z},
\mbox{ where }
\ds Z= \sum_I e^{-{E_I/T}}$$
(where $I,J$ can be single indices or multiple indices),
one may check that, after applying
the definition \req{HSh} or \req{HNSh}, eq. \req{HlnZ} holds.
Therefore,
we might as well have started our calculation of $H$ from
eq. \req{HlnZ} itself (and we will do so for the $N$-dimensional case).
Analogously,
$-T \, \ln(Z)$ plays the role of the Helmholtz free energy $F$,
satisfying the relation $\ds S=-{dF\over dT}$.

A `finely' or thinly spaced discretization means that $\lambda$
should be small. However, the above expression of $\theta(\beta; m)$
as a series is obviously inadequate when 
$\ds\beta= {\lambda^2\over {2\pi}} \ll 1 $.
Such a difficulty will be overcome by recalling the remarkable
theta function identity (see e.g. ref.\cite{Bate})
\beq
\theta(\beta; 0)=
{1 \over \sqrt{\beta}} \, \theta\left( {1 \over \beta} ; 0 \right) .
\label{JI}
\eeq
Applying now this identity 
to \req{Hlnth} or \req{HlnZ},
expanding each part for small $\lambda$ and differentiating, one finds
\beq
\begin{array}{ll}
H= {1 \over \ln(2)}&\ds\left[ {1 \over 2} + \ln\left( \sqrt{ 2\pi} 
\over \lambda \right)
+2 \left( 1-{2 \pi^2 \over \lambda^2} \right) e^{-{2 \pi^2 \over \lambda^2 }}
-2  e^{-{4 \pi^2 \over \lambda^2 }} + {8 \over 3}  e^{-{6 \pi^2 \over
 \lambda^2 }}
+{\cal O}\left( 
{2 \pi \over \lambda^2}  e^{-{8 \pi^2 \over \lambda^2}} 
\right) \right]
\end{array}
\label{H1app}
\eeq
which, regarded as an 
expansion, is quickly convergent  for
$0 < \lambda \ll \sqrt{2\pi}$. (One should notice that, actually, the two 
expressions have a generous overlap around $\beta \simeq 1$ where both 
converge and any of them can be consistently used).

There is no explicit dependence on $\sigma$, as the only relevant variable
is the relative discretization size $\lambda$.
Even for moderately large values of $\lambda$, the next-to-leading
part of $H$ is very small: e.g. for
$\lambda=1$ we have
${e^{-{\pi\over\beta}} \over \beta}=
2\pi \, {e^{-{2 \pi^2 \over \lambda^2}} \over \lambda^2} \simeq
1.7 \cdot 10^{-8}$,
${e^{-2{\pi\over\beta}} \over \beta}=
2\pi \, {e^{-{4 \pi^2 \over \lambda^2}} \over \lambda^2} \simeq
4.5 \cdot 10^{-17}$;
at $\lambda=1/2$ these two quantities become
$1.3 \cdot 10^{-33}$ and $6.6 \cdot 10^{-68}$, respectively.
Neglecting such terms,
we easily obtain 
a good approximate formula, which may be reexpressed as
\beq
\begin{array}{lll}
H&=&\ds\log_2 \left( \sqrt{ 2 \pi e \over \lambda^2} \right)
+{\cal O}\left( 
{2 \pi \over \lambda^2} \, e^{-{2 \pi^2 \over \lambda^2}}, \dots
\right) ,
\end{array}
\label{H12lapp}
\eeq
with $\lambda$ given by \req{deflaapp}. This yields eq. \req{H12l}

\subsection{$N$-dimensional case \label{AppNd}}
After looking at the $Z$ of eq. \req{ZN1},
let's introduce, for convenience, the new notations
\beq
\sigma \equiv \mbox{min}( \{ \sigma_1, \dots, \sigma_N \} ), \hspace{1cm}
\chi^{-1} \equiv \sigma^2 \, C^{-1} , \hspace{1cm}
\lambda \equiv { \Delta\eta \over \sigma } ,
\label{nnot}
\eeq
which enable us to write
\beq
Z= \sum_{n_1, \dots, n_N=-\infty}^{\infty}
e^{\ds -{\lambda^2 \over 2}
(n_1, \dots, n_N)^T \, {\chi}^{-1} \, (n_1, \dots, n_N) } .
\eeq
In terms of the multidimensional Jacobi theta function
\beq
\theta_N(\beta | M) \equiv  \sum_{n_1, \dots, n_N= -\infty}^{\infty}
e^{\ds -\pi \beta \, \sum_{i,j=1}^N M_{ij} n_i n_j }  ,
\eeq
we can put
\beq
Z= \theta_N\left( \beta \vert {\chi}^{-1} \right) ,
\hspace{1cm} \beta\equiv {\lambda^2 \over 2\pi} \equiv{1 \over T} .
\label{defbN}
\eeq
By \req{HlnZ}, the joint Shannon entropy (now $H\equiv H_N$) becomes
\beq
H= {d \over dT}\left[ T \log_2(Z) \right]
= \log_2\left[ \theta_N\left( \beta \vert {\chi}^{-1} \right) \right]
-\beta{d \over d\beta} 
\log_2\left[ \theta_N\left( \beta \vert {\chi}^{-1} \right) \right] .
\eeq
We are interested in approximations for small $\beta$, but the present expressions
are inadequate for this situation.
The way out is to take advantage of a Jacobi identity
for multidimensional theta functions, namely,
\beq
\theta_N\left( \beta \vert M \right)
={1 \over [\Det(M)]^{1/2} \beta^{N/2} } \,
\theta_N\left( \left. {1 \over \beta} \right\vert M^* \right) ,
\eeq
which, unlike the initial expression, may be expanded for small $\beta$.
Doing so
(and noting that $C$ has to be real when viewed in configuration space),
\beq
\begin{array}{lll}
H&=&\ds {N \over \ln(2) }\left[
{1 \over 2} + \ln\left( 1\over\sqrt{\beta} \right)
+{1 \over 2N} \, \ln \Det({\chi})
+{\cal O}\left( {1  \over \beta} e^{ -{\pi\over\beta}\mbox{min}(C/\sigma^2)}
 \right)
\right] \\
&=&\ds \log_2\left[
\sqrt{ \Det\left( {2 \pi e \over (\Delta\eta)^2} \, {C} \right) }
\right]
+{\cal O}\left( 
N \, {2\pi\sigma^2 \over (\Delta\eta)^2 } e^{-{2\pi^2 \mbox{\ssz min}(C)
\over(\Delta\eta)^2}} 
\right) ,
\end{array}
\label{HlogDetapp}
\eeq
where min($C$) means the minimum over the (positive) eigenvalues
 of the correlation matrix $C$, and where the relations \req{nnot} 
and the definition
of $\beta$ in \req{defbN} have been used. More terms of this expansion
can be obtained explicitly by using eq.\req{H1app}. Notice, however, that
each order of \req{H1app} gives rise here, in principle,
 to $N$ different orders
(the first $N$ of them corresponding to the sequence of eigenvalues of
$C$, increasing in magnitude).  The bottom line gives us
\req{HlogDet}.

\section{Appendix: useful Fourier-space results \label{secFS}}

\subsection{Discrete Fourier transforms \label{AppDFT}}
The continuous transforms taken as reference are
\beq
\left.
\begin{array}{lll}
\widehat{\eta}(k) &=&\ds \int dx \, e^{-i kx} \, \eta(x) , \\
\eta(x) &=&\ds \int {dk \over 2\pi} \, e^{i kx} \, \widehat{\eta}(k) . \\
\end{array}
\right\} ,
\label{contFTpair}
\eeq
where $k$ and $x$ are a pair of conjugate variables.
Discretizing them,
\beq
\left.
\begin{array}{lll}
k_n &=& n \Delta k , \\
x_n &=& n \Delta x ,
\end{array}
\right\}
\label{discrxk}
\eeq
and calling
\beq
\left.
\begin{array}{lll}
\widehat{\eta}_n &\equiv& \widehat{\eta}( k_n ) , \\
\eta_n &\equiv&\eta( x_n )  ,
\end{array}
\right\}
\eeq
we construct discrete transforms which, in the continuum limit, reproduce
\req{contFTpair}:
\beq
\left.
\begin{array}{lll}
\widehat{\eta}_n &=&\ds \Delta x \sum_m  e^{-i k_n x_m} \, \eta_m , \\
\eta_n &=&\ds {\Delta k \over 2\pi} \sum_m e^{i k_m x_n} \, \widehat{\eta}_m .
\end{array}
\right\}
\eeq
Taking into account \req{discrxk}
and the correct relation between sampling intervals, i.e.
$\ds\Delta k= {2\pi \over N \Delta x}$, one realizes that
\beq 
k_n \, x_m = k_m \, x_n = {2 \pi \over N} m n.
\eeq
Therefore, we can write
\beq
\left.
\begin{array}{llll}
\widehat{\eta}_n &=&\ds \Delta x&\ds (W \, \vec{\eta})_n , \\
\eta_n &=&\ds {\Delta k \over 2\pi}&\ds( W^* \, \widehat{\eta})_n .
\end{array}
\right\}
\eeq
where $W$ is the symmetric matrix with coefficients 
$W_{mn}= e^{\ds i {2\pi \over N} mn}$. 
After renaming 
\beq
\begin{array}{lll}
x&\to & t ,\\
k& \to &\omega ,
\end{array}
\eeq
this yields the expressions \req{DFTpair}.

\subsection{Fourier-space relation involving $\Det(C)$ \label{subsecDetC}}
For convenience, we prefer to handle
the Fourier-space representation of $C$ 
---which we shall denote by $\widehat{C}$---
rather than $C$ itself (we will see that $\widehat{C}$ is simpler).
A vector $\vec\eta$ and its discrete Fourier transform $\widehat{\eta}$
are related by expressions of the type 
\beq
\left.
\begin{array}{lll}
\ds\widehat{\eta}&=&\ds \Delta t \, W \vec\eta ,  \\
\ds\vec\eta&=&\ds {\Delta\omega \over 2 \pi} \, W^* \widehat{\eta} ,
\end{array}
\right\}
\label{DFTpair}
\eeq
where $W$ indicates a matrix whose coefficients are given by
$W_{mn}= e^{\ds i \, {2 \pi \over N} \, m n}$
(see subsec. \ref{AppDFT}).
$\Delta t$ is a $t$-interval which now has to
be interpreted as the time lapse between two successive
Fourier `samplings'. If we imagine that $\eta_j=\eta(t_j)$,
then $t_j-t_{j-1}= \Delta t$, $\forall j$.
$\Delta\omega$ is the corresponding interval in `angular frequency'
or conjugate space.
Taking into account the usual
relation between the sampling interval and the associated
angular frequency (or conjugate momentum) range that can
be correctly sampled in conjugate space, one has the following
relation between $\Delta t$, $\Delta\omega$ and $N$:
\beq
\begin{array}{lll}
\Delta\omega &=&\ds 2\pi \, {1 \over N \Delta t} ,
\end{array}
\label{dxdk}
\eeq
The discrete values of $\omega$ are
\beq
\omega_j= j \Delta\omega, \ j=-N/2, \dots, N/2.
\label{omegaj}
\eeq
Let $\omega_{\mbox{\ssz min}}$ and $\omega_{\mbox{\ssz Max}}$
denote the minimum and maximum nonzero absolute values of $\omega$. Then,
\beq
\begin{array}{rclcrcrcr}
\omega_{\mbox{\ssz min}}&\equiv&
\omega_1&=&\Delta\omega &=&\ds{2 \pi \over N\Delta t}&=& 2\pi f_{\mbox{\ssz min}} , \\
\omega_{\mbox{\ssz Max}}&\equiv&
\omega_{N/2}&=&\ds {N \over 2} \Delta\omega &=&\ds {\pi \over \Delta t}&=& 2\pi f_{\mbox{\ssz Max}} .
\end{array}
\label{kMkm}
\eeq
We have here introduced frequencies ---$f$'s--- in the way $\omega=2\pi f$,
as usual.

Furthermore, by the form of its coefficients and by eq.\req{DFTpair},
it is clear that the $W$ matrix satisfies
\beq
\begin{array}{lll}
W^T &=& W, \\
W^{-1}&=&\ds {\Delta t \Delta\omega \over 2 \pi} \, W^* ,
\end{array}
\label{WTWm1}
\eeq
and, consequently,
$\ds
W^{-1} = {\Delta t \Delta\omega \over 2 \pi}  W^{T \, *}.
$
In other words, up to a multiplicative scalar constant,
$W$ is a unitary operator.
Taking now formula \req{defC}, we
apply \req{DFTpair} and \req{WTWm1} to write the $C$ matrix
in terms of Fourier-space objects, and quickly obtain
\beq
C= 
\left( { \Delta\omega \over 2\pi \, \Delta t } \right) 
W^{-1} \, \widehat{C} \, W ,
\label{ChC}
\eeq
where $\widehat{C}$ is the above mentioned Fourier-space representation
of $C$, i.e., it is the matrix whose coefficients read
\beq
\widehat{C}_{jk} = \langle \widehat{\eta}_j \widehat{\eta}_k^* \rangle .
\eeq
Formula \req{ChC} is telling us that
\beq 
\Det(C) = 
\left( { \Delta\omega \over 2 \pi \, \Delta t } \right)^N 
\Det\left( \widehat{C} \right)
\label{DetChC-app}
\eeq
independently of $W$.

\subsection{Power Spectrum \label{subsecPS}}

Recall first the definition \req{defC} of the covariance matrix:
$\ds C_{jk}= \langle \eta_j \eta_k^* \rangle $,
where $\langle \dots \rangle$ denotes statistical average over realizations
of the stochastic process $\eta$.
For a stationary stochastic process we have that
$C_{jk}=C_{j-k}$ can only be a function of $j-k$, eg 
the covariance matrix is Toeplitz matrix.
It is a simple exercise to show that in this case the covariance matrix in 
Fourier space $\widehat{C}$ is always diagonal:
\beq
\ds \widehat{C}_{jk} \equiv \langle \widehat\eta_j \widehat\eta_k^* \rangle 
\propto \delta_{jk}
\eeq
The {\it power spectrum} is then defined as:
\beq
\ds \widehat{C}_{jk} \equiv P(\omega_j) \, 
{\delta_{jk}\over{\Delta\omega}} ,
\label{CP-app}
\eeq
in analogy with the continuous definition:
\beq
\langle \widehat\eta(\omega) \, \widehat\eta^*(\omega') \rangle =
P(\omega) \, \delta_{\mbox{\ssz Dirac}}(\omega-\omega').
\eeq

\subsection{Nonzero mean \label{Appnzm}}
In the practical handling of data, it is sometimes necessary to introduce
offsets, with the consequence that a variable which had initially
a zero mean may lose such a property. Assuming that 
$\langle \eta \rangle =0$, let's suppose that an offset $a\in{\bf R}$ is added
to $\eta$. Thus, for the new variable $\eta'\equiv\eta+a$ one has
$\langle \eta' \rangle =a$. From the definition 
\req{defC}, we find that the covariance matrix of $\eta'$ is just
\beq
C'= C+ a^2 I,
\eeq
where $C$ is the covariance matrix of $\eta$, and $I$ is the identity matrix.
Relating now covariance and power spectrum 
by eqs. \req{ChC} and \req{CP} ---or \req{CP-app}---,
one realizes that the new power spectrum is simply
\beq
P'= P+ 2 \pi \, \Delta t \, a^2 I,
\label{PprimeP}
\eeq
i.e., the previous one shifted by a constant, which corresponds to the
entropy change coming from the 
knowlegde of $\langle \eta' \rangle =a$. Procceding in this way, 
it is possible to use the same computational methods as for the zero-mean case,
with the only difference that $P$ has to be modified according to 
eq.\req{PprimeP}.

\section{Appendix: The continuous random variable case \label{appcrv}}
As it is well-known, Shannon's entropy was firstly designed to deal with
discrete random variables
\begin{equation}
H(\eta)\equiv -\sum_{j}p(\eta_j)\log_2[p(\eta_j)] ,
\end{equation}
where the index runs through all possible different countable values of the r.v..
The problem with the continuous r.v. is that different $\eta_j$'s 
do not form a partition.
To define $H(\eta)$ we form first the discrete r.v. $\eta_{\Delta}$
obtained by rounding off $\eta$
\begin{equation} 
\eta_{\Delta} \equiv n\Delta\eta\quad,
\mbox{if}\quad n\Delta\eta-\Delta\eta< x\leq n\Delta\eta .
\end{equation}
Clearly,      
\begin{equation}
P(\eta_{\Delta}=n\Delta\eta)=P(n\Delta\eta-\Delta\eta< \eta \leq n\Delta\eta)
=\int_{n\Delta\eta-\Delta\eta}^{n\Delta\eta} d\eta \, f(\eta)=\Delta\eta \, \bar{f}(n\Delta\eta) ,
\end{equation}
where $\bar{f}(n\Delta\eta)$ is a number between the maximum and minimum of $f(\eta)$ 
in the interval $(n\Delta\eta-\Delta\eta, n\Delta\eta)$.
Applying Shannon's definition, we have:
\begin{equation}
H(\eta_{\Delta})=
-\sum_{n=-\infty}^{\infty}\Delta\eta \, \bar{f}(n\Delta\eta)\log_2[\Delta\eta \, \bar{f}(n\Delta\eta)]
\end{equation}\\ 
and, since
\begin{equation}                                                        
\sum_{n=-\infty}^{\infty}\Delta\eta \, \bar{f}(n\Delta\eta) =
\int_{-\infty}^{\infty}d\eta \, f(\eta)=1,
\end{equation}\\
we conclude that 
\begin{equation}
H(\eta_{\Delta})=
-\log_2(\Delta\eta)
-\sum_{n=-\infty}^{\infty}\Delta\eta \, \bar{f}(n\Delta\eta)\log_2[ \bar{f}(n\Delta\eta) ] .
\end{equation}\\
As $\Delta\eta\to 0$, the r.v. $\eta_{\Delta}$ tends to $\eta$; however,
its entropy $H(\eta_{\Delta})$ tends to $\infty$ because
$-\log_2\Delta\eta\to\infty$.
This is why we define the entropy $H(\eta)$ of $\eta$ not as the limit of
$H(\eta_{\Delta})$ but as the limit of the sum
$H(\eta_{\Delta})+\log_2(\Delta\eta)$ when $\Delta\eta \to 0$, i.e.:
\begin{equation}
H(\eta_{\Delta})
+\log_2(\Delta\eta) 
\longrightarrow
\int_{-\infty}^{\infty} d\eta \, f(\eta)\log_2[ f(\eta) ]\quad
\mbox{as}\quad \Delta\eta \to 0.    
\end{equation}
So, the definition of `entropy' for a continuous variable is:
\begin{equation} 
H(\eta)=\int_{-\infty}^{\infty} d\eta \, f(\eta)\log_2[ f(\eta) ] ,
\end{equation}\\
where the integration extends only over the region where $f(\eta)\neq 0$,
as we have $f(\eta)\log_2[ f(\eta) ]=0$ if $f(\eta)=0$.
This `entropy' is more usually called {\it differential entropy} in the literature
and its definition can also be extended to multivariate probability distributions. 
It is easy to see then that the above limit translates into:
\begin{equation} 
H(\vec{\eta}_{\Delta})+\log_2(\Delta\eta)^N \longrightarrow
\int_{-\infty}^{\infty} d\vec{\eta} \,  f(\vec{\eta})\log_2[ f(\vec{\eta}) ] \quad
\mbox{as} \quad \Delta\eta \to 0
\end{equation}
when the $N$-dimensional space of $\vec\eta$ is latticed with $\Delta\eta$-boxes.
So, we could approximate
$H(\vec{\eta}_{\Delta}) \simeq -\log_2(\Delta\eta)^N + H(\vec{\eta})$.
In our case we have defined the compression ratio as
$\ds c_{\mbox{\ssz r, opt}} \equiv\frac{\mbox{average length}}{h}$, where
$\ds h\equiv\frac{H(\vec{\eta}_{\Delta})}{N}$.
If we look just back to the approximate:
\begin{equation}
h\approx -\log_2(\Delta\eta) + H(\vec{\eta})/N.
\end{equation}
The last summand in the previous expression is the average uncertainty per sample 
in a block of $N$ consecutive samples. 
The limit $N \to \infty$ of it is what is known as {\it differential entropy 
rate}:

\begin{equation}
\ol{h}(\vec{\eta})=\lim_{N \to \infty}\frac{H(\eta_1, \dots, \eta_N)}{N}
\end{equation}

So, if we imagine that we have a stochastic process infinitely long
and $\vec{\eta}$ is a vector r.v. whose dimension tends to infinity
(i.e. $\eta_j=\eta(t_j)$ and we take samples for a long time or just many samples)
we could then approximate:

\begin{equation} 
h \simeq -\log_2(\Delta\eta) + \ol{h}(\vec{\eta}).
\label{hhcont}
\end{equation}

Regarding $\ol{h}(\vec{\eta})$ as the `continuous part'
of the entropy per component ---i.e. $h_{\mbox{\ssz cont}}$---,
this relation amounts to eq. \req{hhcontn0}.

\subsection{Entropy in the continuous case\label{AppExa}}

For the one-dimensional Gaussian distribution in eq.\req{defpj} 
it is straight
forward to show that:
\beq
\ol{h} \equiv  h_{\mbox{\ssz cont}} = \log_2 \left[ \sqrt{ 2 \pi e} \, \sigma \right],
\eeq
in agreement with eq.\req{H12l} in the limit of small $\Delta \eta$, as expected
from the comments in the previous section.
For the case of N-dimensional Gaussian noise with correlations,
we can use the fact that $\ol{h}(\vec{\eta})$ is well-known 
(see eg. \cite{Papou})
for a Gaussian stochastic process with power spectrum $P(\omega)$:
\begin{equation} 
\label{hcont}
\ol{h}(\vec{\eta})=
\log_2[\sqrt{2\pi e}]+\frac{1}{4\pi}\int_{-\pi}^{\pi} d\omega \, \log_2[\tilde P(\omega)] .
\end{equation}
where $\tilde P(\omega)$ refers to the discrete stochastic 
process derived from the continuous one $P(\omega)$ 
by the relation
\begin{equation}
\tilde P(\omega)=\frac{1}{\Delta t} \sum_{m=-\infty}^{\infty}
P\left(\frac{\omega+2\pi m}{\Delta t}\right) , 
\qquad -\pi \leq \omega \leq \pi ,
\label{PPprimesumm}
\end{equation}
where $\Delta t$ 
is the sampling interval that discretizes the process.
For power spectra with a bandwidth limitation this reduces to
(see \cite{Ih}):
\begin{equation}
\tilde P(\omega)=\frac{1}{\Delta t} P(\frac{1}{\Delta t} \omega) 
\label{PPprime}
\end{equation}
where $\tilde P(\omega)$ refers to the process 
$\eta_n=\eta( t= n\Delta t )$.
In this case we can do a simple change of variables $\omega'=\omega/\Delta t$ in
eq.\req{hcont} to find:
\beq
\label{hcont2}
\frac{1}{4\pi}\int_{-\pi}^{\pi} d\omega \, \log_2[\tilde P(\omega)]
= -2 \log_2{\Delta t} + {\Delta t \over{2\pi}} \int_0^{\pi/\Delta t}
 d\omega' \, \log_2[P(\omega')].
\eeq
where we have used the parity of $P(\omega)$ and the fact that the range
in eq.\req{hcont} is symmetric .
Recalling that $\omega_{\mbox{\ssz Max}}=\pi/\Delta t$ 
and we are using $\omega_{\mbox{\ssz min}} \simeq 0$ we see can that this
calculation is equivalent to the Euler-Maclaurin
summation formula eq.\req{Eulersf}, so that the continous calculation
of the entropy given by eq.\req{hcont} and eq.\req{hhcont}
yields identical results to those of the discrete
calculation eq.\req{defhcorr} in the limit of large $N$.

\vspace*{1cm}
\ni {\bf\Large Acknowledgements}
\vspace*{3mm}

\ni 
 We would like to thank Pablo Fosalba
for stimulating disscusions.
This work has been supported by CSIC, and by DGES (MEC), 
project PB96-0925.

\end{document}